\documentclass[10pt]{article}

\usepackage{booktabs}
\usepackage{times}
\usepackage[ruled,resetcount,linesnumbered,noend]{algorithm2e}
\usepackage{graphicx}
\usepackage{amsmath}
\usepackage{paralist}
\usepackage{multirow}
\usepackage{caption}
\usepackage{courier}
\usepackage{subcaption}
\usepackage{amssymb}
\usepackage{balance}
\usepackage{bookmark}
\usepackage{microtype}
\usepackage{comment}
\usepackage{paralist}
\usepackage{etoolbox}

\newtheorem{lemma}{\textbf{Lemma}}
\newtheorem{theorem}{\textbf{Theorem}}

\SetKw{Event}{upon}
\SetKwBlock{Do}{}{}
\SetKw{For}{for}
\SetKw{In}{in}
\SetKw{Proc}{procedure}
\SetKw{Before}{before}
\SetKw{After}{after}
\SetKw{Around}{around}
\SetKw{Funct}{function}
\SetKw{Ret}{return}
\SetKwBlock{Atomic}{atomic}{}

\newcommand{\spara}[1]{\vspace*{0.05in}\noindent\textbf{#1.}}

\newenvironment {squishlist}
{\begin{list}{$\bullet$}
  { \setlength{\itemsep}{0pt}
     \setlength{\parsep}{3pt}
     \setlength{\topsep}{3pt}
     \setlength{\partopsep}{0pt}
     \setlength{\leftmargin}{1.5em}
     \setlength{\labelwidth}{1em}
     \setlength{\labelsep}{0.5em} } }
{\end{list}}

\newcommand{\name}{Eli\'a\xspace}
\newcommand{\NAME}{Eli\'a\xspace}
\newcommand{\rubis}{RUBiS\xspace}
\newcommand{\tpcw}{TPC-W\xspace}
\newcommand{\mysql}{MySQL\xspace}
\newcommand{\mysqlcluster}{MySQL Cluster\xspace}
\newcommand{\namealgo}{operation Partitioning\xspace}
\newcommand{\Namealgo}{Operation Partitioning\xspace}
\newcommand{\NameAlgo}{Operation Partitioning\xspace}

\newcommand{\NameProt}{Conveyor Belt\xspace}
\newcommand{\code}[1]{\texttt{#1}\xspace}

\begin{document}

\title{Scaling Out ACID Applications with \NameAlgo}

\author{Habib Saissi, Marco Serafini$\dagger$, Neeraj Suri\\ \\
\{\code{saissi}, \code{suri}\}\code{@cs.tu-darmstadt.de}\\
Technische Universit\"at Darmstadt, Germany\\
$\dagger$\code{mserafini@qf.org.qa}\\
Qatar Computing Research Institute, Qatar
}
\date{}
\maketitle
\thispagestyle{empty}

\begin{abstract}
OLTP applications with high workloads that cannot be served by a single server need to scale out to multiple servers.
Typically, scaling out entails assigning a different partition of the application state to each server.
But data partitioning is at odds with preserving the strong consistency guarantees of ACID transactions, a fundamental building block of many OLTP applications.
The more we scale out and spread data across multiple servers, the more frequent distributed transactions accessing data at different servers will be.
With a large number of servers, the high cost of distributed transactions makes scaling out ineffective or even detrimental.

In this paper we propose {\em \NameAlgo}, a novel paradigm to scale out OLTP applications that require ACID guarantees.
\NameAlgo indirectly partitions data across servers by partitioning the application's {\em operations}  through static analysis.
This partitioning of operations yields to a lock-free {\em \NameProt} protocol for distributed coordination, which can scale out {\em unmodified} applications running on top of {\em unmodified} database management systems.
We implement the protocol in a system called \name\ and use it to scale out two applications, \tpcw\ and \rubis. 
Our experiments show that \name\ can increase maximum throughput by up to 4.2x and reduce latency by up to 58.6x compared to \mysqlcluster while at the same time providing a stronger isolation guarantee (serializability instead of read committed).
\end{abstract}


\section{Introduction}
\label{sec:intro}

Online transaction processing (OLTP) applications, such as online shopping services, bidding services, or social networking systems, need to scale in order to handle demanding workloads.
One common way to increase capacity is to run the application on top of multiple servers, a process that is called {\em scale out}.
These applications often use ACID transactions with {\em strong consistency guarantees}, which give the impression of being executed in some sequential order even if they are executed concurrently.

It is well known that strong consistency guarantees substantially simplify the design of applications, but make scaling out challenging.
A common approach to scale out is {\em data partitioning}, which partitions the persistent state of the application across multiple servers.
If a transaction needs to access data across multiple partitions, it is executed as a {\em distributed transaction}, which requires coordination across multiple servers.
Distributed transactions are costly and represent the main bottleneck hindering scale out.
The more servers we use, the more frequent distributed transactions become.
As such, there is a bound on the degree of scale out that can be achieved with ACID applications.
For example, our evaluation shows that the \tpcw benchmark on MySQL Cluster reaches its peak performance with four servers, after which adding more servers is not beneficial anymore.

Scaling out efficiently entails solving two problems: finding a good way to partition data, and finding an efficient algorithm to keep servers consistent.
In this paper we introduce the concept of \NameAlgo, a novel approach to address these two problems in an integrated manner.

\NameAlgo takes an indirect approach to data partitioning.
It maps each client operation to a specific server responsible for executing it, trying to associate conflicting operations to the same server whenever possible.
This partitioning of the operations yields, indirectly, to a (partial) partitioning of the data.
By focusing on partitioning operations rather than data, \NameAlgo makes it possible to perform partitioning based only on {\em static analysis} of the application code.
This analysis is entirely automated, unlike existing data partitioning approaches that require human expertise and/or running samples of a workload in order to come up with good partitions (e.g.~\cite{schism,horticulture}).
In addition, the analysis can be applied to {\em unmodified} application code, without the need for the user to provide additional information about the semantic of the application.

\NameAlgo not only makes partitioning easier, it also enables designing a more efficient coordination algorithm, called {\em \NameProt} protocol, that guarantees serializability across multiple servers~\cite{serializability}.
The protocol obviates one of the main sources of inefficiency of distributed transactions: holding locks at multiple servers until a transaction is completed.
\NameProt is a {\em lock-free} protocol, which critically relies on the {\em operation classification} produced by the static analysis of the application code.
Servers use a token passing scheme to execute ``global" operations that, according to the classification, require coordination with other servers.
When a server receives a global operation, it simply puts it on hold until it receives the token, without impairing the progress of other ``local" operations that require no coordination.
Once a server gets the token, global operations are executed efficiently in a batch.
Our evaluation shows that the performance of \NameAlgo  is superior to data partitioning with distributed transactions, both in terms of performance with a given number of servers and in terms of maximum number of servers that can be effectively utilized.

Compared to recent techniques to speed up distributed transactions, such as Calvin~\cite{calvin}, Lynx~\cite{lynx}, Rococo~\cite{rococo}, Callas~\cite{callas}, and others~\cite{faleiro2015rethinking,transaction-chopping,callas2}, the \NameProt protocol has two main advantages.
First, existing techniques require additional information about the semantic of the application, which must be provided by the user and might not be trivially available, or might not be available at all in some application.
In addition, they require extending the application to provide this information and/or modifying the application code (e.g. to chop transactions).
The \NameProt protocol does not require any knowledge about the semantic of the application, as it only relies on the automatic \NameAlgo process.
This means that the \NameProt protocol can be used to scale out {\em unmodified} applications.
Second, these techniques must be implemented by designing a new database management or key-value store system.
The \NameProt protocol, by contrast, operates on top of {\em unmodified} single-server database management systems (DBMSs) providing ACID transactions.
Using an unmodified DBMS, without requiring any specific low-level support for distributed transactions, makes it easier to run \NameProt on top of a wide range of technologies as a middleware.

To show the practical viability of our approach, this paper presents \name, a new middleware to scale out Java applications (Web applications running on Apache Tomcat in our use cases) and unmodified JDBC-compatible databases (MySQL in our use cases).  
We used \name to scale out two common OLTP benchmarks, \tpcw and \rubis. 
In a LAN setup, where all servers are running within one datacenter, \name increases maximum throughput by 4.2x and decreases minimal latency by 58.6x compared to \mysqlcluster, a prototypical system based on data partitioning. 
This is particularly remarkable if we consider that \name is not only faster but also provides a significantly stronger consistency guarantee (serializability instead of read committed isolation, which is the only isolation level offered by \mysqlcluster). 
In a WAN (i.e., geographically distributed) setup, scaling out using \name reduces latency by up to 47.9x and increases throughput by up to 2.91x compared to a centralized setting. 

Overall, we make the following contributions:
\begin{compactitem}
\item We introduce \NameAlgo, a scale out solution for OLTP applications that requires ACID transactions. \NameAlgo is the first approach to use automated static analysis to indirectly partition data;
\item present the \NameProt protocol, an efficient lock-free coordination algorithm that relies on the operation classification produced by \NameAlgo;
\item implement \name, a middleware that uses \NameAlgo\ to scale out unmodified DBMSs with ACID transactions;
\item use \name to scale out \tpcw and \rubis. In a LAN setting, \name outperforms \mysqlcluster by 4.2x in terms of throughput and 58.6x in terms of latency. In a WAN setting, \name improves throughput and latency by up to 2.9x and 47.9x respectively.
\end{compactitem}


\section{Overview}

\NameAlgo considers the problem of improving the throughput and latency of an ACID application running on top of a DBMS by scaling out, i.e., running instances of the DBMS on top of multiple servers.
These DBMS instances are kept consistent by running the \NameProt protocol on top of them.
The protocol coordinates the execution of operations and guarantees serializability.
We now give an overview of the steps required by \NameAlgo.

\spara{Offline static analysis} The \NameAlgo process consists of three main steps, which are separate but intertwined. 
First, an {\em automated partitioning} step is performed to determine how to partition operations.
\NameAlgo requires that the code of the application is known a priori.
This is a sound assumption for many Web and enterprise OLTP applications, since they typically run a fixed set of transactions. 
The partitioning algorithm statically analyzes read-write conflicts between operations to minimize cross partition conflicts.
Partitioning avoids coordination by routing conflicting operations to the same server as much as possible.
Operations that have no conflicts with operations at other servers can be executed locally and immediately, without coordination with other servers.
In particular, partitioning tries to minimize the type of conflicts that require coordination in the \NameProt protocol.
We describe the automated partitioning algorithm in Section~\ref{sec:partition}.

Next, the {\em operation classification} step uses the partitioning obtained in the previous step to classify operations as commutative, local, or global, based on the amount of coordination they require.
Commutative and local operations can be executed immediately without distributed coordination, unlike global operations.
Operations classification, which is also an automated process, is described in Section~\ref{sec:classes}.

\spara{Online scale-out algorithm} The previous two steps of offline analysis produce a partitioning criteria and an operation classification.
These are taken as input by the {\em \NameProt protocol}, which runs the application on multiple servers and ensures consistency.
The protocol is described in Section~\ref{sec:protocol}.

The protocol is implemented by \name, a scale out middleware that integrates with unmodified applications and interfaces with unmodified external DBMSs.
We describe the technical details of this integration in Section~\ref{sec:system}.

\section{Operation Partitioning}
\label{sec:classification}

We start by describing the first two steps in our approach: automated partitioning algorithm and operations classification.

\spara{Application code: transactions vs. operations}
We consider applications keeping all their persistent state in a DBMS.
The application code consists of a set of \textbf{transactions} that modify the state of the DBMS.
Transactions are expressed as procedures having a certain number of input parameters.
For example, a transaction could be the procedure \texttt{createCart(sid)}, which creates a shopping cart with id \texttt{sid}.
An \textbf{operation} corresponds to a request to execute the transaction with a set of concrete values for its input parameters.
For example, a client operations can invoke the operation \texttt{createCart(5)} to create a cart with id 5.


\spara{Operation conflicts}
The application state is stored by the DBMS, and logically consists of a set of variables (i.e., tuples).
A state assignment (or simply state) $S$ assigns a value to each variable accessed by the application.
Let $O$ be the set of all possible operations that can be executed by the application.
The read set $R(o)$ of an operation $o \in O$ consists of all variables that $o$ may read when it executes on any state $S$.
Similarly, the write set $W(o)$ of $o$ is the set of all variables that $o$ may write to if it executes on any state $S$.
Two operation $o_1$ and $o_2$ have a {\em write conflict} if their write sets intersect, i.e., $W(o_1) \cap W(o_2) \neq \emptyset$.
We say that $o_1$ {\em reads from} $o_2$ if $R(o_1) \cap W(o_2) \neq \emptyset$.
In either cases, we say that $o_1$ and $o_2$ conflict with each other.

\subsection{Automatic Partitioning}
\label{sec:partition}

The automatic partitioning step generates a partitioning of operations that minimizes conflicts. We now describe how we automate this process. 

To identify operation conflicts we need to specify read and write sets of the operations. 
First, we show how to extract and express read and write sets from the source code.
Next, we describe the automated partitioning algorithm, which takes read and write sets as input and determines an {\em operation partitioning array} $P$.
The operation partition array associates every transaction $t$ to one of its input parameters.
This {\em partitioning parameter} is used by the \NameProt  protocol to route every operation $o$ of type $t$ to a server.
After an operation partitioning array $P$ is determined, classifying operations is straightforward and automatic as we will see.



%

\spara{Extracting read/write sets}
An OLTP application usually has a relatively small number of transaction, which can correspond to a huge number of possible operations.
Therefore, the \namealgo algorithm operates at the level of granularity of transactions, and for each transaction determines a read and a write set.
These sets are determined in a static and pessimistic way: they include all variables that could be accessed in any execution performed against any database state.
An entry $e$ in either sets is a pair $e=\langle A, C\rangle$, where $A$ is a set of {\em accessed attributes} and $C$ is a {\em condition}.

The accessed attributes set in the read set contains all table attributes (i.e., columns) that are read and returned as output of the transaction.
In the write set, it contains all table attributes that are updated by the transaction.
The condition of a read or write set is the predicate used to select the specific rows in the table for which the attributes are modified.

Read and write sets are generic concepts, but we now give a concrete example based on the type of applications we targeted in this work.
These applications consist of a set of transactions that access a database through SQL queries.
Consider for example the \code{doCart} transaction in the \tpcw\ benchmark, which updates a shopping cart with id \code{sid} by adding, removing or updating item with id \code{iid} in a quantity \code{q}.
The pseudocode of the transaction is the following:
\begin{small}
\begin{verbatim}
    doCart(sid, iid, q){
	        ...
	        exec("UPDATE SHOPPING_CARTS 
	        SET QTY = q WHERE ID = sid 
	        AND I_ID = iid");
	        ...	
    }
\end{verbatim}
\end{small}
\noindent \Namealgo extract reads and write sets by looking at {\em all} SQL statements contained in the transaction, regardless of the execution path.
While conservative, this approach has proven good enough for our purpose.
We used Java parser \cite{jparser} to extract SQL queries and to map input parameters to the used query parameters.

With this information at hand, we can define read and write sets.
Each SQL statement corresponds to an entry in one of the sets.
Consider for example the SQL statement highlighted in the pseudocode and rename the table \code{SHOPPING\_CARTS} as \code{SC} for brevity.
This statement corresponds to a write set entry $e$.
The accessed attribute for $e$ is specified in the \code{UPDATE} clause, so $e.A=$\code{SC.QTY}.
Insert SQL query also correspond to entries in the write set and their accessed attribute is specified in the \code{INSERT} statement, while for read set entries the accessed attribute corresponds to the \code{SELECT} query.
The condition of the entry corresponds to the content of the \code{WHERE} clause of the query, so in this case $e.C=(\code{SC.ID = sid}$ $\wedge$ $\code{SC.I\_ID = iid})$.
The condition binds the value of the input parameters of the transaction, which are \code{sid} and \code{iid} in this case, with the values of the table attributes of the specific rows for which the attributes in $e.A$ are accessed by the transaction, \code{SC.ID} and \code{SC.I\_ID = iid} in our example.

\begin{algorithm}[t]
	\caption{Partitioning Algorithm.}
	\label{alg:classification-new}
	\begin{footnotesize}
		\SetKwInOut{Input}{input}\SetKwInOut{Output}{output}
		\Input{ Set $T$ of transactions}
		\Input{ Read set $R_t$ and write set $W_t$ for each transaction $t$}
		\Output{ Array $P$ of partitioning parameters $P[t]$ for each transaction $t$}
		
		\BlankLine

		\tcp{Conflict detection}		
		\ForEach{pair $t, t' \in T$}{
			$C_{t,t'} \leftarrow  \textit{false}$\;
			\If{$\exists r \in R_t, w \in W_{t'}: r.A \cap w.A \neq \emptyset$}{
				$C_{t,t'} \leftarrow C_{t,t'} \vee (r.C \wedge w.C)$\;
			}
			\If{$\exists w \in W_t, r \in R_{t'}: w.A \cap r.A \neq \emptyset$}{
				$C_{t,t'} \leftarrow C_{t,t'} \vee (w.C \wedge r.C)$\;
			}
			\If{$\exists w \in W_t, w' \in W_{t'}: w.A \cap w'.A \neq \emptyset$}{
				$C_{t,t'} \leftarrow C_{t,t'} \vee (w.C \wedge w'.C)$\;
			}
			\If{$C_{t,t'}$ is satisfiable}{
				\textit{Conflicts} $\leftarrow \textit{Conflicts} \cup C_{t,t'}$\;
			}
		}
		
		\tcp{Partitioning optimization}	
		\Return $\min_P$ cost$(P, \textit{Conflicts})$\; 
		
		\BlankLine
		
		\tcp{Estimate the volume of conflicts}	
		\Funct cost($P$, \textit{Conflicts}) \Do{
			\ForEach{$C_{t,t'} \in $ Conflicts}{
				$k \leftarrow P[t]$\;
				$k' \leftarrow P[t']$\;
				\ForEach{table attribute $A$}{
					remove all clauses $(k = A \wedge k' = A \wedge \ldots)$ from $C_{t,t'}$\;
				}
				\If {$C_{t,t'}$ not satisfiable}{
					remove $C_{t,t'}$ from \textit{Conflicts}\;
				}
			}
			\Return $\sum_{C_{t,t'} \in \textit{Conflicts}} \textit{weight}(t) + \textit{weight}(t')$\;
		}
		
	\end{footnotesize}
\end{algorithm}

\spara{Conflict detection phase}
The partitioning algorithm is illustrated in Algorithm~\ref{alg:classification-new}.
The first phase of the algorithm is {\em conflict detection}, which looks at all pairs of transactions that have a conflict on some table attribute.
A conflict between transactions occurs if some of the operations relative to these transaction can conflict, according to the definition of Section~\ref{sec:classes}.
For each pair of transactions $(t,t')$, it builds a condition predicate $C_{t,t'}$, in disjunctive normal form, that expresses the condition that the values of the input parameters of $t$ and $t'$ must take so that a conflict occurs on the same row(s) of the same table(s).
In other words, the condition characterizes the set operations of the two transactions that are conflicting.
If a conflict between the two transactions is possible, $C_{t,t'}$ is added to a set called {\em Conflicts}.
Note that we also consider self-conflicts, that is, conflicts between two operations of the same transactions where $t = t'$.

Let us consider again the TPC-W example. 
The \code{createCart} transaction creates a new row in the \code{SHOPPING\_CARTS} table (again renamed \code{SC} for brevity) such that \code{SC.ID = sid}, where \code{sid} is the id of the shopping cart and is an input parameter of \code{createCart}: 
\begin{small}
	\begin{verbatim}
    createCart(sid){
       ...
       exec("INSERT INTO SHOPPING_CARTS 
       (ID) VALUES (sid)");
       ...
    }
	\end{verbatim}
\end{small}
\noindent The write set of \code{createCart} contains entry $e = \langle$ \code{SC.ID}$, $\code{SC.ID = sid} $\rangle$.
Given the write set of \code{doCart}, we derive that there is a write-write conflict between the two transactions with condition $C_{t,t'}$: 
\begin{equation*}
\begin{small}
\label{eqn:conflict}
(\code{SC.ID = sid}) \wedge (\code{SC.ID = sid'}) \wedge (\code{SC.I\_ID = iid'})
\end{small}
\end{equation*}
where \code{sid} is a parameter of \code{createCart} and \code{sid'} and \code{iid'} are parameters of \code{doCart}.

\spara{Partitioning optimization phase}
The next phase is called {\em partitioning optimization} and it finds the operation partitioning array $P$ that minimizes global operations, as defined in Section~\ref{sec:classes}.
The partitioning can reduce the cost of conflicts by mapping two conflicting operation to the same partition, and thus server, such that the conflict becomes local.

The cost function finds out the potential an operation partitioning has to eliminate conflicts.
Consider two transactions $t$ and $t'$ that conflict, and let $k$ and $k'$ be the parameters used for their partitioning.
\Namealgo uses the same deterministic routing function for all operations, so two operations with the same value of their partitioning parameters $k$ and $k'$ will be sent to the same server.
Therefore, all conflicts that arise because of a necessary condition $k = k'$ will be local to one server, and they will not require global coordination.
The most common case when this condition arises is when $k$ and $k'$ are used to identify a row based on the value of the same attribute $A$, so there is a clause in the conflict condition of the form: $(k = A \wedge k' = A \wedge \ldots)$

Let us revisit again our running TPC-W example and let $P$ be an operation partitioning array such that \code{sid} is the partitioning parameter for both \code{doCart} and \code{createCart} transactions.
The conflict condition in the previous equation
is of the form $(k = A \wedge k' = A \wedge \ldots)$, where $k=$\code{sid}, $k'=$\code{sid'}, and $A =$\code{SC.ID}.
This condition is equivalent to saying that the conflict among the two transactions arises only if \code{sid}$=$\code{sid'}.
As the same deterministic routing function is used for both transactions, conflicting operations will always be sent to the same server. This means that we can remove this conflict from the {\em Conflicts} set. 

After removing all conflicts that become local thanks to an operation partitioning array $P$, we can estimate the cost of the remaining global conflicts by summing up the weight of the conflicting transactions in {\em Conflicts}.
If we assign to each transaction a weight of $1$, the algorithm will minimize the number of conflicting transactions.
If an estimate of the relative frequency of the transaction is known, it can be used as a weight to improve cost estimation.

The algorithm searches for the operation partitioning array that minimizes the cost. 
In the workloads we considered, and in most practical transactional workloads, the number of transaction types and their parameters is not very large, so an exhaustive search of all possible partitionings to find the best one is feasible.
However, the algorithm can also use of more sophisticated search strategies.

\spara{Multiple partitioning parameters}
The full algorithm also considers multiple partitioning attributes by looking at each  parameter independently to find a partition.
If in all cases the resulting partition is the same, we consider the operation local and send it to that partition.
Otherwise, it is not possible to map the operation to one partition and it is marked as global.

\spara{Applicability of the algorithm}
Although our static analysis tool targets transactional applications using SQL statements, Algorithm~\ref{alg:classification-new} is generic and can be applied to other types of applications.
For example, a key-value store can be seen as a single table with two attributes.
In our implementation, however, we target application code using basic SQL queries.
For partitioning, we require that potential partitioning parameters are involved in \code{WHERE} clauses only in atomic conditions in an equality form.
The rest of the clause can contain arbitrary conditions. 
Parameters used in atomic conditions that are not in equality form are ignored for partitioning, and other alternatives are tried out.
We also do not consider complex SQL constructs such as nested queries and triggers.
\subsection{Classes of Operations}
\label{sec:classes}
With a partitioning of operations at hand, we can now describe the operation classification logic.

\Namealgo identifies three classes of operations: {\em commutative}, {\em local}, and {\em global}.
Commutative operations can be executed by any server, and do not require prior coordination.
These operations do not conflict with any other operation.
Local operations are partitioned, so they need to be executed by a specific server, but they do not require prior coordination.
Even though a local operation $l$ can have conflicts, no operation operations executed at a different server than the one assigned to $l$ depends on the effect of executing $l$. Both commutative and local operations are not replicated.
Finally, global operations require coordination before they are executed and are replicated.

%

\spara{Commutative operations}
The first step of classification is to identify the operations that do not have conflicts with any other operation in $O$.
We call these operations {\em commutative operations}.
A commutative operation is either a read-only operation accessing immutable state, for example an operation reading some fixed configuration parameter, or a write-only operations whose writes are never read by any other operations, for example a logging operation.
Commutative operations do not require any coordination: when a server receives a commutative operation, it can execute it locally and send a reply without any synchronization.
We denote with $C \subseteq O$ the subset of all commutative operations in $O$.

\spara{Local and Global operations}
Consider now the set $O \setminus C$ of operations that have some conflict with some other operations.
We classify these operations as local or global by first partitioning them and by assigning each partition to a different server in the system.
We then classify each operation as follows.
An operation $o$ is a {\em local operation} if: (i) $o$ does not have a write conflict with any other operation in a different partition, and (ii) no other operation from a different partition reads from $o$.
We denote with $L_p$ the set of local operations in the partition assigned to a server $p$.
A local operation $l$ associated to a specific server can be executed immediately at that server without any prior coordination.
In fact, it follows from conditions (i) and (ii) that no other operation associated with another server depends on the effects of $l$.

All operations that are neither commutative nor local are called {\em global operations}. 
We denote with $G_p$ the set of global operations in the partition assigned to server $p$.
Since executing global operations entails coordination among servers in \NameProt, it is important to find an operation partitioning that minimizes them.
Note that global operations are also assigned to partitions, and are therefore only executed by a dedicated server, because they may read from other local operations which are only seen by that server. 
Allowing global operations to execute on arbitrary servers might results in them attempting to read unavailable data.

\begin{figure}
\centering
\includegraphics[width=\columnwidth]{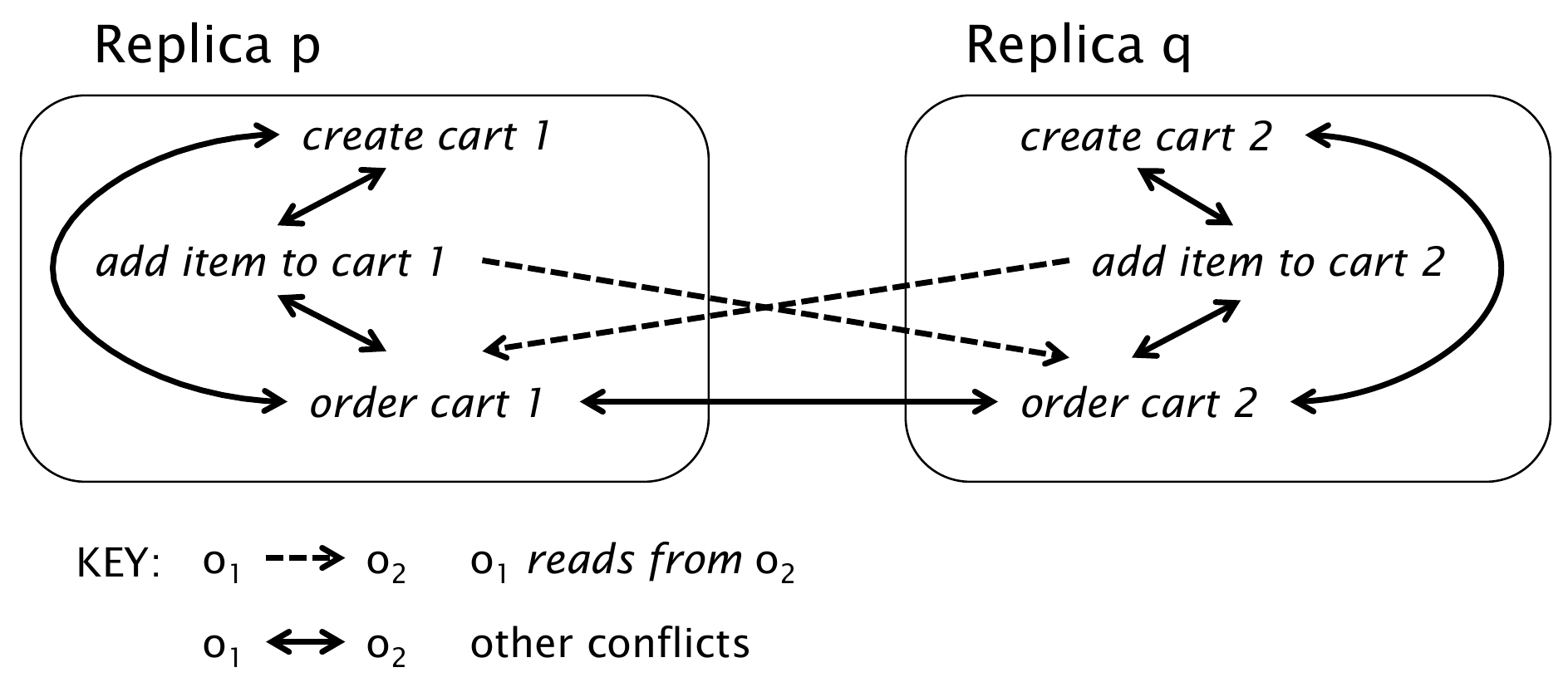} 
\caption{A classification of operations in the online store example. The {\em order} operation is global, the other operations are local.} 
\label{fig:conflicts}
\end{figure}

\spara{Example}
Consider the example of an online store application whose operations must be classified.
The application has transactions that allow clients to create a cart, 
add and remove items, 
and eventually proceed to checkout. 
Each cart is assigned a unique id.
For each cart id $c$ we have the following three operations: {\em create} a cart $c$, {\em add} a quantity $a$ of items of type $t$ to $c$ provided that there are sufficient items on stock, and finally {\em order} all items currently present in the cart $c$.

Assume that operations are partitioned based on the value of the cart id $c$.
The conflicts among operations in two sample partitions are illustrated in Figure~\ref{fig:conflicts}.
For example, operations that add an item to a cart do that only if the item is available in the stock.
The stock level of an item can be modified by order operations, which remove elements from it.
Therefore, add operations on cart $c$ read-from order operations for the same cart, and order operations have write conflicts with other order operations on different carts.

\Namealgo\ classifies operations according to the partitioning and their conflicts as follows.
Order operations are {\em global} because they have write-write conflicts with operations in other partitions, and because add operations read-from them. 
All other operations are {\em local} because they either have no conflicts with operations at other servers (e.g., create cart operations) or they only read-from from remote operations (e.g., add to cart operations). 

\spara{Comparison with data partitioning}
\NameAlgo can be seen as a mechanism to generate partial data partitions.
The union of write sets of every local operation in a partition corresponds to a data partition.
The subset of the write set of a global operation that is part of cross partition conflicts constitute the set of items that have to be replicated.
For instance, in Figure~\ref{fig:conflicts}, there are two data partitions each containing a cart.
Every cart is assigned to the partition of the operations manipulating it.
Since the order operation, a global operation, writes to an item that is also written to by order operations from other partitions, the item has to be replicated.

Conversely, it is possible to get an \NameAlgo from a data partitioning scheme.
Given a data partitioning function $f$ that maps every data entry to a partition, we generate an operation partitioning by ensuring that every local operation has no entries in its conflict set belonging to two distinct partitions. More formally, an operation $o$ is local if for every conflicting operation $o'$, for every variables $x, y \in W(o) \cap W(o')$ or $W(o) \cap R(o')$, we have $f(x) = f(y)$.
In that case, the server responsible for $o$ also maintains the partition $f(x)$.

\section{The \NameProt Protocol}
\label{sec:protocol}

We now describe \NameProt, a lock-free scale-out coordination algorithm that runs applications on multiple servers to increase their performance compared to a single-server configuration.
The protocol considers applications where \NameAlgo has already been applied to classify the application's operations.
The classification allows a server to immediately execute and reply to as many local operations as possible without coordinating with other servers.
\NameProt implements {\em serializability}~\cite{serializability}, i.e, all clients observe the same sequential execution order of operations.
We provide a correctness proof of the protocol in the appendix.

\begin{algorithm}[t]
\caption{The \NameProt algorithm for server $p$}
\label{alg:coord-passive}
\begin{footnotesize}

\Event receive $\langle \textrm{REQ}, o, c \rangle$ msg from client $c$ where \nllabel{ln:c-begin}\Do{
	\If{$o \in C \cup L_p$\nllabel{ln:cl-begin}}{
		$r, *\leftarrow$ execute($o$)\;
		send $\langle \textrm{REPLY}, r \rangle$ msg to $c$\nllabel{ln:cl-end}\;	
	} \ElseIf{$o \in G_p$\nllabel{ln:g-begin}}{
		append $\langle o,c\rangle$ to $Q$\nllabel{ln:g-end}\;
	} \Else {
		$q \leftarrow$ replica such that $o \in L_q \cup G_q$\nllabel{ln:map-begin}\; 
		send $\langle \textrm{MAP}, q \rangle$ msg to $c$\nllabel{ln:map-end}\;
	}
}

\BlankLine

\Event event \textsc{receiveToken}$(T)$ \nllabel{ln:primary-begin}\Do{
	\ForEach{$\langle u,q \rangle \in T$\nllabel{ln:token-start}}{
		\If{$p = q$}{
      remove $\langle u,q \rangle$ from $T$\nllabel{ln:token-remove}\;
		}
		\Else{
			apply($u$)\nllabel{ln:token-end}\;
		}
	}
	$Q' \leftarrow $ atomic-snapshot($Q$)\nllabel{ln:copy}\;
	\ForEach{$\langle o,c \rangle \in Q'$\nllabel{ln:q-start}}{
		$r, u \leftarrow$ execute($o$)\nllabel{ln:q-execute}\;
    append $\langle u, p \rangle$ to $T$\nllabel{ln:q-append}\;
		send $\langle \textrm{REPLY}, r \rangle$ msg to $c$\;
		remove $\langle o,c \rangle$ from $Q$\nllabel{ln:q-end}\;
	}
	\textsc{passToken}$(T)$\nllabel{ln:token-release}\;
} 
\end{footnotesize}
\end{algorithm}

\spara{Preliminaries}
We start by clarifying the functioning of the application running with \NameProt.
The application is ran by an event-driven multi-threaded server.
Whenever a event, such as the receiving of a client request, is triggered, the server dispatches its handling to a thread.
When a server receives requests (\textsc{REQ} message) from clients, the assigned thread executes the request and sends a (\textsc{REPLY} message) back or a redirect message (\textsc{MAP} message) if the client contacted the wrong server.

\NameProt orchestrates the execution of the application by invoking its request execution logic. 
In the pseudocode, we abstract this logic as an $\emph{execute}(o)$ function, which executes an operation $o$ and produces a reply $r$, along with a state update $u$ that we describe shortly.
We consider applications that store their state on a database management system (DBMS).
Each server runs a local stand-alone DBMS instance, i.e., instances of the DBMS at different servers do not communicate with each other.
When an application executes an operation, it accesses its local state by invoking a sequence of database queries on the local DBMS instance.
All queries invoked by the same operation are enclosed within a single database transaction.
We assume that the DBMS can execute transactions in parallel and guarantee serializability.

The state update $u$ returned by $\emph{execute}(o)$ is the update-only query that includes all updates to the database generated during the execution of $o$.
Extracting $u$ is one of the features of the \name system, which we detail in Section~\ref{sec:system}.
When value of the state update is irrelevant, we it this with a star.

\NameProt executes each operation only once, at a single server. 
It replicates the effects of operations executed at other servers simply by directly applying the corresponding state update onto the local DBMS instance.
This is denoted in the pseudocode by the function $\emph{apply}(u)$.

The  algorithm requires executing operation partitioning and classification as preliminary steps before it is started.
These steps partition operations among the sets $\{C, L_1, \ldots, L_N, G_1, \ldots, G_N\}$, where $N$ is the number of servers in the system, $C$ contains commutative operations, and $L_p$ and $G_p$ contain the local and global operations, respectively, assigned to server $p$.

\spara{Handling commutative and local operations}
The pseudocode of the \NameProt algorithm for a server $p$ is shown in Algorithm~\ref{alg:coord-passive}.
The algorithm handles operations differently based on their classifications.
Commutative and local operations are executed locally and a reply is immediately sent back to the client without coordination (Lines~\ref{ln:cl-begin}-\ref{ln:cl-end}).

It is easy to see why commutative operations do not require coordination. 
For local operations the argument is more subtle.
As discussed in Section~\ref{sec:classes}, the updates made by local operations are not directly read by any other operation running at other servers.
However, these updates might indirectly impact remote operations transitively, my means of a global operation.
Consider again the cart example of Section~\ref{sec:classes}.
If a (local) {\em add} operation adds an item $i$ to a cart and a subsequent (global) {\em order} operation places an order for that cart, the result is that the stock of item $i$ is reduced, and this impacts operations at all servers.
The \NameProt protocol in this case only propagates the state update of the {\em order} operation, which includes the reduction in the stock of item $i$.
This is sufficient to ensure serializability, and there is no need to propagate the fact that the item was previously added to the cart.
The protocol can thus avoid propagating local operation thanks to its use of state updates, or in other words, of passive instead of active replication~\cite{difference}:
propagating and executing operations at all servers would require \NameProt to propagate the {\em add} operation as well in order to guarantee a consistent execution of the {\em order} operation at all servers.
we refer the reader to the proof of correctness in the appendix.

\spara{Handling global operations}
Global operations require coordination among servers to agree on a total order of execution.
\NameProt uses a token based scheme.
The token is passed around in a predefined order to ensure that global operations are totally ordered.
At any time, only the server holding the token, also called the primary, is allowed to execute global operations.
Otherwise, the server appends the operations to a queue $Q$ for execution at a later time.
Note that the queue $Q$ must be thread-safe since Algorithm~\ref{alg:coord-passive} is multi-threaded.


\spara{Handling the token}
Upon receiving the token $T$, server $p$ invokes one \textsc{recieveToken}$(T)$ event at $p$, becoming the primary server.
Like any other event, a specific \textsc{recieveToken}$(T)$ event is handled in isolation by a single thread, while multiple other threads might be concurrently handling client requests.
The token contains a sequence of tuples $\langle u, q \rangle$ where $u$ is the update of a global operation that has previously been executed at some server $q$.
As soon as a server becomes primary, it applies all the updates in the token that are from other servers and removes its own updates as they have been already applied at all other servers (Lines~\ref{ln:token-start}-\ref{ln:token-end}).
Next, the primary needs to execute the global operations that have been enqueued locally into $Q$.
In order to ensure liveness, the primary copies an atomic snapshot $Q'$ of the $Q$ queue containing global operations submitted to $p$ that have been waiting for execution (Line~\ref{ln:copy}).
This is because $Q$ is concurrently modified by multiple threads.
Without copying an atomic snapshot, $p$ might stay stuck executing incoming global operations in $Q$ that are constantly being appended by other threads, and never give up the token. 
Then, $p$ iterates over all global operations that have been pending up to that point in the $Q'$ queue (Lines~\ref{ln:q-start}-\ref{ln:q-end}). 
The server executes each operation $o$ in the queue, sends a reply $r$ to the client $c$ and appends the resulting update $u$ to the token before removing the operations that have been appended.

For efficiency reasons, our actual implementation of the \NameProt protocol executes the operations in $Q'$ in parallel.
Consequently, the DBMS executes multiple concurrent transactions generated by these operations.
\name must be able to extract the logical serial order in which the DBMS executes these concurrent transactions, since this serial order must be the same as the order in which the corresponding state updates are added to the token and thus applied at other servers.
Section~\ref{sec:system} describes how \name achieves this.
Finally, the server gives up the primary role by calling \textsc{passToken}$(T)$ to pass the token to the next server (Line~\ref{ln:token-release}).

\spara{Redirections}
If clients know how the operations are partitioned, they send their operations directly to the server responsible for it.
This should be the common case as it is for our example applications.
However, if clients send an operation to the wrong server, the server will reply with the identity of the server responsible for the operation (Lines~\ref{ln:map-begin}-\ref{ln:map-end}).

\spara{Fault Tolerance}
The \NameProt protocol considers replication for fault tolerance as a complementary and orthogonal issue.
More precisely, the protocol assumes that each of the servers can tolerate faults and there is no message loss.
Making a server fault tolerant is an orthogonal issue: for example, a Paxos group could implement the abstraction of a logical fault tolerant server.
For message loss, the token can simply be passed using a reliable channel among fault-tolerant servers.


\section{The \NAME\ System}
\label{sec:system}

We have developed \name, a middleware that uses the \NameProt protocol at its core to ensure coordination-freedom of local operations.
\name\ supports multi-threaded applications, where concurrent threads execute operations on a shared application state.
Each server stores the application state in a local DBMS offering  serializable transactions. We implemented \name in Java and it consists of about 2k lines of code.

\begin{figure}
\centering
\includegraphics[width=\columnwidth]{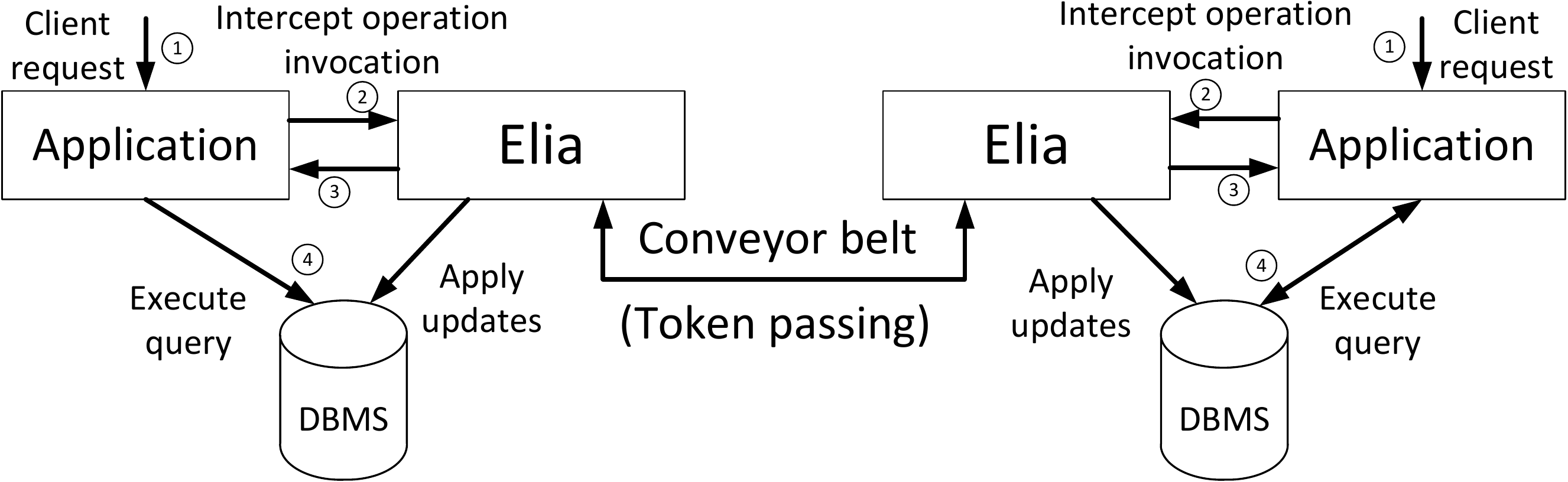} 
\caption{\NAME system overview in a deployment with two servers. The numbered arrows indicate the execution flow of operations that are mapped to the server. Global operations are held after step 2 until the server acquires the token. The \NameProt protocol propagates state updates for global operations executed at other servers. \name directly applies these state updates onto the DBMS.}
\label{fig:env}
\end{figure}

\spara{Overview}
In our implementation the mutli-threaded application is a web application with a pool of threads to handle incoming HTTP requests.
\NAME works by intercepting the interaction between the application threads and the DBMS.

A key design choice in \NAME was that we wanted to scale out unmodified applications running on top of an unmodified external DBMSs.
In particular, the DBMS is seen as a black box by \name.
Our current implementation of \NAME intercepts JDBC interactions between application threads and the DBMS, so we can support any data store offering a JDBC interface interacting with a serializable DBMS.
For example, in our evaluation we use (unmodified) MySQL as underlying data store.
This makes \name\ easier to adopt and allows dealing with fault tolerance at the data store level.
Furthermore, the application does not have to be manually modified to work together with \NAME but is instead automatically instrumented.

In Figure \ref{fig:env}, we show an overview of a multi-threaded server application running with \NAME.
\NAME interacts directly with the application, and the other servers.
Before an application starts executing an operation, it invokes \NAME and waits until it is allowed to proceed according to the \NameProt protocol.
When it is time to execute the operation, \NAME gives back control to the application to resume.
Next, the application executes an instrumented version of the operation that allows the recording of the resulting updates.
The application interacts directly with the DBMS and delivers the operation updates in the same order of their execution back to \NAME.
Additionally, \NAME runs a separate module that deals with token passing and applies received updates directly to the database.

In the description of the \NameProt in Section~\ref{sec:protocol}, we left out a few implementation details about the interaction between \name, the application server, and the underlying DBMS, namely: how to extract state updates, how to manage and parallelize the execution of global operations, and how to trace the sequential execution order of global operations.
We describe these details in the following.


\spara{Extracting state updates}
First, we describe how \name extracts a state update from the execution of an operation.
\name does this by intercepting the execution flow of operations in the application.
We do this by automatically instrumenting the application code to enable the interaction with \NAME.
An operation can execute multiple accesses to the underlying DBMS.
Before the first access, the operation starts a transaction, which is terminated when the operation terminates.
This ensures that the local execution of concurrent operations is equivalent to a sequential execution on the DBMS.
The interaction between the application and \NAME is completely transparent. 
In order to obtain the state updates produced by global operations, \NAME records changes to the DBMS state by intercepting interactions between the application and the DBMS, which occur through JDBC.
Every time the application begins executing a global operation, our instrumentation generates an operation object that is used to store the state updates.
\NAME then uses the operation object as a wrapper to JDBC: every time the application invokes a statement $s$  mutating the state (e.g., \code{UPDATE}), it does so through the operations object instead of JDBC. 
The operation object appends $s$ to the sequence of SQL query statements invoked within the operation and then passes $s$ to JDBC.
At the end of the transaction, the sequence of SQL statements in the operation object represents the sequence of state mutations that can be executed by other servers to reproduce the operation, that is the update that has to be passed to the other servers.

\spara{Parallelizing the execution of global operations}
We now describe how our implementation of the \NameProt protocol in \name handles global operations.
The handling of local and commutative operations is identical to the description in Algorithm \ref{alg:coord-passive}, but \name optimizes the handling of global operations by executing them in parallel.
As discussed in Section~\ref{sec:protocol}, there is a single thread that handles the event of receiving a token.
We call this thread the {\em token thread}.
We call the other threads that handle global operations, and are  waiting for the permission of executing them, the {\em handling threads}.
In Algorithm~\ref{alg:coord-passive} the handling thread appends a global operation to $Q$ and returns, leaving the actual execution of the operation to the token thread.
This would require the handling thread to store a copy of the HTTP request handling context necessary to reply to the client and make it available to the token thread.
As this would induce a substantial overhead, we opted for having the handling thread wait for the server to receive the token before executing the operation with the necessary HTTP request context.
Concretely, we extend the queue $Q$ to contain an initially locked lock for every pending operations that the handling thread attempts to acquire and goes to sleep until it is unlocked by the token thread.
When the token thread executes, it iterates over the pending operations and notifies the sleeping threads to proceed and execute the operation.
The token thread then blocks until all pending operations finished execution using a semaphore initialized with the number of pending operations.
Once an operation finishes execution, it adds its update the token queue and decreases the counter of the semaphore.
When the token thread is awakened again it releases the token and returns.
This implementation has the additional benefit of speeding up the execution of global operations as they are handled concurrently by multiple threads.
Since global operations are executed concurrently, it is important that the order in which the transaction updates are added to the token correspond to the order in which they were executed by the database. 

\spara{Tracing the sequential order of global operations}
Next, we show how we ensure that the execution order of global operations matches their order in the token sequence of updates.
\NAME assumes that the DBMS provides serializability, so it executes transactions in a total order.
\NAME must record the serial execution order of the database to make sure that the state updates are broadcast consistently with this order.
To this end, the wrapper operation objects uses a reference queue $U$ to the token to capture the order of state updates.

In our implementation we assume that transactions ensure serializability using pessimistic locking: before a transaction accesses a data item $i$, the transaction acquires a lock and releases it only after the transaction is committed or aborted.
When the application requires a transaction $t$ for operation $o$ to commit, \NAME intercepts this call, appends to $U$ the state update $u_o$ produced by $o$, and then invokes the commit.
Since the DBMS uses pessimistic locking, \NAME knows that $t$ has already taken locks on all the data items it accesses before it invokes the commit.
Therefore, any concurrent transaction $t'$ for an operation $o'$ that has a conflict with $t$ will not be able to invoke commit until $t$ has committed and released its locks.
The thread executing $t'$ will thus append $o'$ to $U$ only after $t$ has finished appending $o$, so the order of the operations in $U$ is consistent with the execution order of $t$ and $t'$.
Updates that do not conflict can be added to $U$ in any order: \NAME uses a concurrent queue implementation to allow safe concurrent updates from multiple threads.

\section{Case Studies}
We present two widely used benchmarks as case studies: \tpcw, an online store system~\cite{tpcw}, and \rubis, an auction website~\cite{rubis}. 
Both benchmarks are implemented in the Java programming language as Java servlets running inside an Apache Tomcat container. 

\begin{table*}[t]
	\centering
	\footnotesize
	\begin{tabular}{|c|c|c|c|c|c|c|c|c|c|c|}
		\hline
		\multirow{2}{*}{Application} & \multicolumn{5}{|c|}{Transaction classification} & \multirow{2}{*}{Total} & \multicolumn{4}{|c|}{Operation frequencies} \\
		\cline{2-6}  \cline{8-11}
		& L & G & C & L/G & Read-only &  & L & G & C &  Read-only \\
		\hline
		\hline
		\tpcw & 10 & 5 & 5 & -- & 13 & 20 & 47\% & 39\% & 14\% & 73\%  \\
		\rubis & 11 & 4 & 3 & 8 & 17 & 26 &64\% & 8\% & 28\% &  85\% \\
		\hline
	\end{tabular}
	\caption{Operations Classification and Frequencies in the used workloads.}
	\label{tab:classification}
\end{table*}
\spara{\tpcw}
\tpcw\ ~\cite{tpcw} is an online bookstore. It handles 14 different client requests such as browsing through books, creating users, adding books to shopping carts or ordering book. The application keeps a persistent state in a database of 10 tables. A \code{SHOPPING\_CARTS} table to store the shopping carts for every user or a \code{ITEMS} table for the available books, among others. 
On average a client request invokes between 2 and 3 operations. 
In total, there are 20 transactions of which 13 are read-only. The rest of the operations either update, delete or insert records in, possibly multiple, tables. 
In \tpcw, \namealgo could identify 10 local, 5 global and 5 commutative transactions  (see Table~\ref{tab:classification}). The local transactions mainly involve updating customer data, and are partitioned by customer id, or manipulations of the shopping carts, and are partitioned by cart id. \name allows generating server-specific unique ids, which guarantee that clients requests partitioned by a given id can be served by the that generated that id. This is important in WAN settings. Global transactions involve ordering books or administrative operations such as updating the books list. The commutative transactions read from immutable tables.

\spara{\rubis}
\rubis\ ~\cite{rubis} is an online auction web application modeled after eBay \cite{ebay}. \rubis defines 20 client requests, such as putting items for sale, viewing personal profiles, bidding or browsing items. The persistent state of the application is stored in 8 tables database. For example, the \code{BIDS} table stores the currents bids and the \code{USERS} the registered bidders.
Similar to \tpcw\ , the requests handlers are not atomic and consist of invocations of multiple operations. 
There are 26 transactions in total of which 17 are read-only. 
In Rubis, \namealgo uses a double-key scheme, whereby many operations are partitioned by both user id and item id.
If both parameters route to the same server, the operation is considered local, otherwise it is considered global.
Such partitioning scheme yields to 11 local, 4 global, 3 commutative, and 8 local/global transactions. 
The local transactions involve the user browsing through his personal profile. Global operations include a global search for items based on some criteria or browsing through a user's own bought items. Commutative operations access immutable tables, such as item categories. Local/global operations involve bidding, buying and selling.

\section{Experiments and Evaluation}
To evaluate our approach we design three set of experiments to answer the following research questions:
\begin{squishlist}
	\item[\textbf{RQ1:}] How does \NameProt (\name) compare to a traditional database that scales out using data partitioning and distributed transactions?
	\item[\textbf{RQ2:}] How does \NAME scale out in a geographically distributed setting?
	\item[\textbf{RQ3:}] What is the minimum fraction of local operations that is sufficient to see performance improvements with \NAME?
\end{squishlist}


\spara{Experimental Setup}
We run our experiments on Amazon EC2 T2 Medium instances (nodes).
Each node has 4 GB of RAM, two virtual cores and is equipped with an Amazon EBS standard SSD with a maximal bandwidth 10000 IOPS. The nodes run Ubuntu Server 14.04 LTS 64, MySQL 5.5.49-0 and Apache Tomcat 7.0.52.

In the LAN experiments, all servers are located in the same site (datacenter) in Germany.
For the WAN (geographically distributed) experiments, we place servers in five different sites to simulate a geographically distributed system. 
The sites are in Germany (G), Japan (J), US east (US), Brazil (B), and Australia (A).
We add these locations in the aforementioned order. For example, a configuration with three locations consists of servers in G, J, and US. 
Table \ref{tab:ping} reports the inter-site latencies among servers. 
The intra-site latency, relevant for the local setup, is in the order of 20 ms. 

We used separate client nodes, which have identical configuration as the servers and are located in the same sites. 
In the WAN setting, we use five client nodes in every configuration, one for each location, and direct requests to the closest server.
We equally distribute client threads across client nodes. 

\spara{Benchmarks}
We use \tpcw and \rubis to evaluate \name.
Both come with multiple workload mixes. 
We use a bidding mix with 15\% write operations for \rubis and the shopping mix with 30\% write operations for \tpcw. 
Table \ref{tab:classification} shows a breakdown of operation frequencies in the workloads for both benchmarks.
Both workloads exhibit a considerable number of local operations that can be leveraged by \name.

\begin{table}
	\centering
	\footnotesize
	\begin{tabular}{|c||c|c|c|c|c|}
		\hline
		Locations & G & J & US & B & A \\
		\hline
		\hline
		Germany (G)	& 	20ms & 253ms & 92ms & 193ms & 314ms     \\
		Japan (J)	&  	  & 	X  & 153ms & 282ms & 188ms    \\
		United States (US)  &     &        & X       &  145ms   &   229ms   \\
		Brazil (B)   &     &        &         &    X       &  322ms \\
		\hline
		
	\end{tabular}
	\caption{Inter-site latencies among server in the WAN setting.}
	\label{tab:ping}
\end{table}
\begin{figure}[t]
	\centering
	\begin{subfigure}[b]{0.47\columnwidth}
		\includegraphics[width=1.1\textwidth]{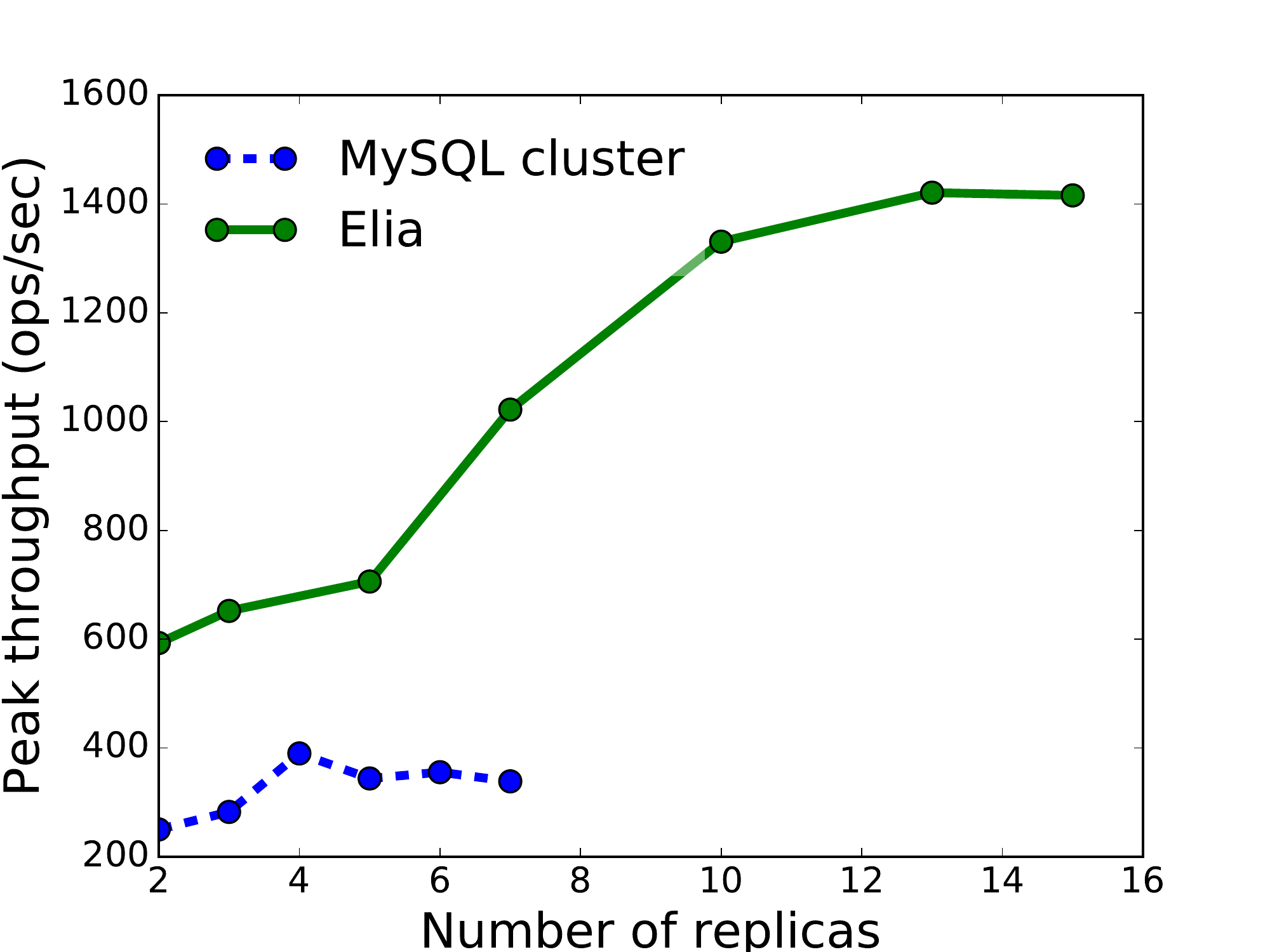}
		\caption{\tpcw.}
		\label{fig:tpcw-gains}
	\end{subfigure}
	~ 
		\begin{subfigure}[b]{0.47\columnwidth}
			\includegraphics[width=1.1\textwidth]{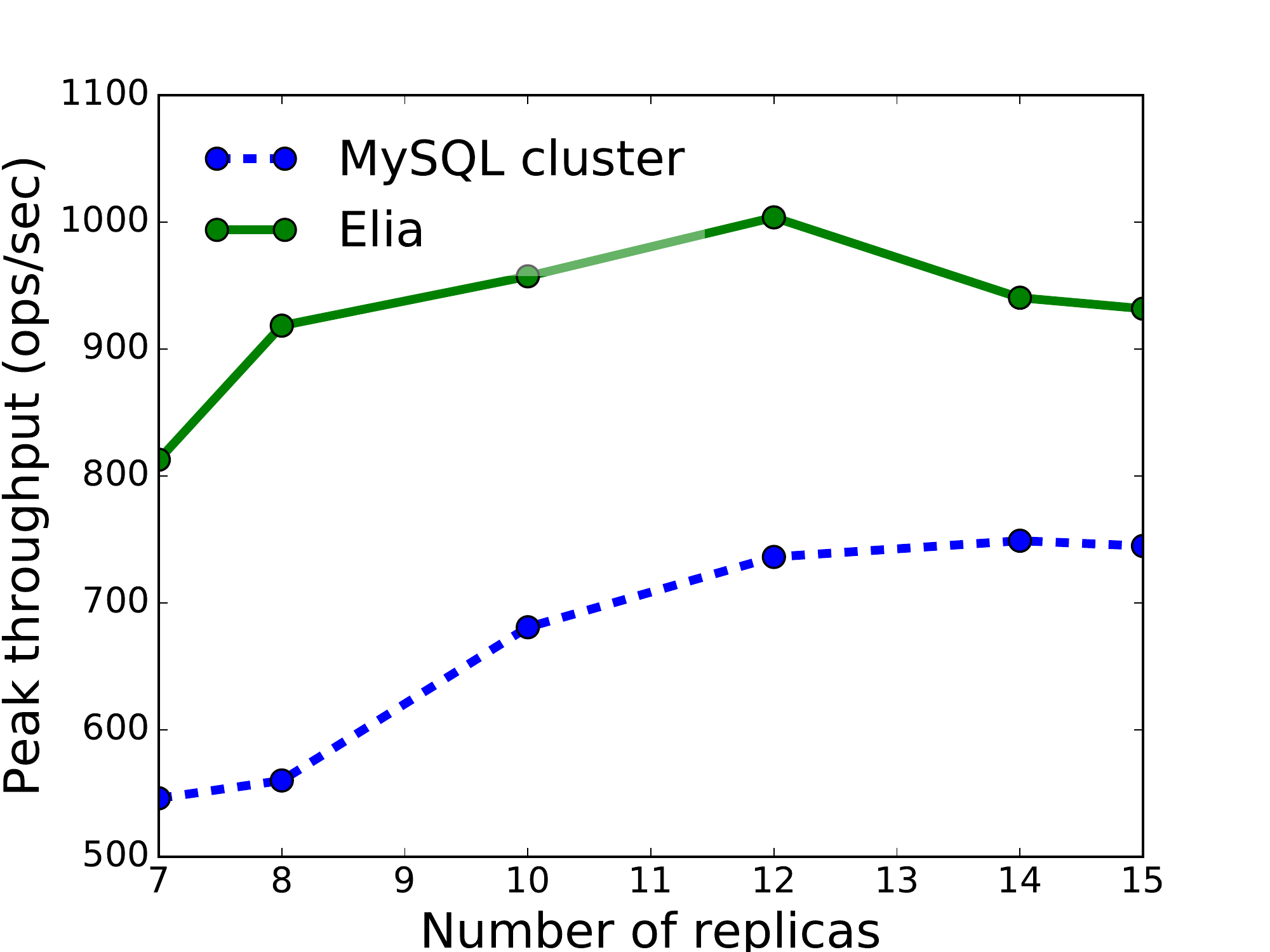}
			\caption{\rubis.}
			\label{fig:rubis-gains}
		\end{subfigure}
	\caption{Scalability of \NAME and \mysqlcluster in a LAN setup.}
	\label{fig:gains}
\end{figure}

\subsection{RQ1: Data Partitioning Comparison} 
In this experiment we compare the performance of \name, against an approach based on data partitioning and distributed transactions.
\mysqlcluster is a version of the popular \mysql DBMS extended with data partitioning capabilities.
It horizontally partitions the database and assigns a partition to each server.
It uses distributed transactions, with pessimistic locking and two-phase commit, for operations that span multiple partitions. 
We choose \mysqlcluster as a baseline because it is a prototypical system combining data partitioning and distributed transactions, and because it is often used as reference for comparison by other state of the art work on distributed transactions like Callas~\cite{callas}.

It is important to note that \mysqlcluster can only provide the {\em read committed} isolation level, whereas \name provides {\em serializability}, which is significantly stronger and more expensive to achieve. 
Nonetheless, \name is still able to achieve a large speedup over \mysqlcluster.

For both benchmarks, we carefully partitioned the database manually using \mysqlcluster.
After running the \Namealgo algorithm, we extracted the resulting data partitioning scheme and applied it to \mysqlcluster.
That is, we use the same data partitioning that result from the \namealgo we apply to the benchmarks. 
For instance, in \tpcw we partition according to customer and cart ids.

We setup each node to serve as \mysqlcluster server and a data node that stores exactly one data partition. We additionally designate one node as the manager for the initial setup.
We use a LAN setting, which is more favorable for \mysqlcluster as distributed transactions are known to perform much better over LANs than over WANs.

\begin{figure}[t]
	\centering
	\begin{subfigure}[b]{0.47\columnwidth}
		\includegraphics[width=1.1\textwidth]{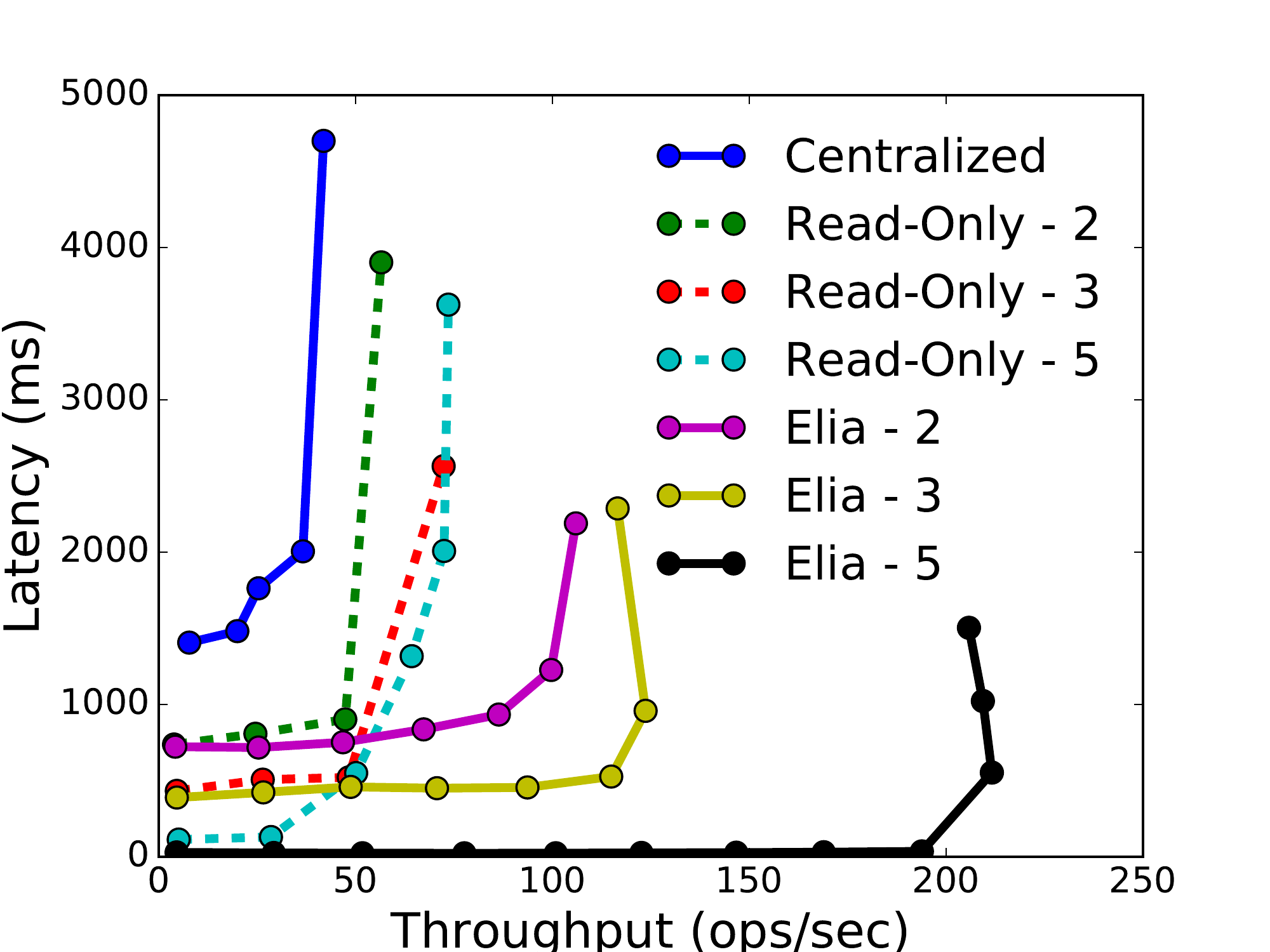}
		\caption{\tpcw.}
		\label{fig:tpcw-wan}
	\end{subfigure}
	~
	\begin{subfigure}[b]{0.47\columnwidth}
		\includegraphics[width=1.1\textwidth]{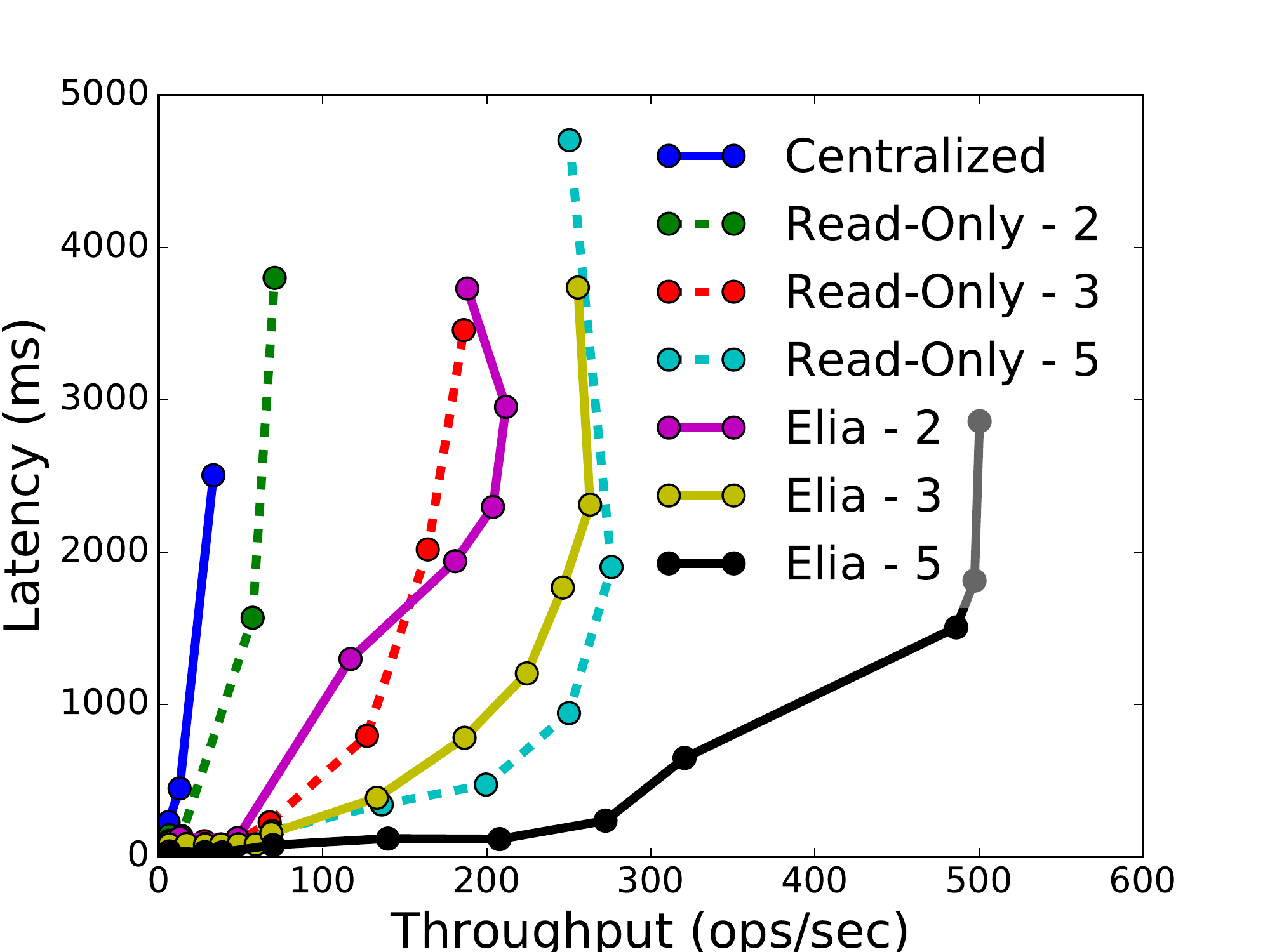}
		\caption{\rubis.}
		\label{fig:rubis-wan}
	\end{subfigure}
	\caption{\NAME vs. baselines in a WAN (geographically distributed) setup.}
	\label{fig:WAN-experiment}
\end{figure}
We examine the scalability of both approaches. In this local setting, we intensify the workload by increasing the number of clients. In Figure \ref{fig:gains} we show how the peak throughput develops while varying the number of servers in the system for \tpcw and \rubis.  
Peak throughput is defined as the maximum throughput a system can sustain while ensuring an average latency of less than 2000ms. 

Figures~\ref{fig:tpcw-gains} and \ref{fig:rubis-gains} show the same trend for both TPC-W and Rubis: as the number of servers grows, the increased cost of distributed coordination eventually outweighs the gain of additional resources to run transactions that require no coordination.
This upper bound in scalabilty represents the inherent cost of achieving strong consistency in the workloads we consider, which are not perfectly partitionable.
Having said that, both figures \ref{fig:tpcw-gains} and \ref{fig:rubis-gains} show that \name scales much better than \mysqlcluster. 
In the case of \tpcw we can see that while the performance of \mysqlcluster starts to degrade with configurations of more than 4 nodes, \name continues to deliver at a much higher throughput until it reaches a configuration of 13 servers. On the other hand, with the \rubis workload, \name and \mysqlcluster reach a point of saturation at the same configuration, namely 12 servers, but still consistently achieves higher throughput.
Overall, \name outperforms \mysqlcluster both in terms of maximal throughput and latency by up to 58.6x for latency and about 4.2x for throughput in the case of \tpcw. 
For \rubis, \name achieves a 1.4 maximal throughput speedup and reduces the latency up to 35.7x.

\name performs significantly better than \mysqlcluster due to the distributed transactions used by the latter to lock rows.
The necessary coordination with remote machines in \mysqlcluster prevents the progress of concurrent conflicting transactions that access the same rows.
In contrast, \name does not lock rows.
When a server receives global operations that require remote coordination, \name merely enqueues the operations until the server gets the token.
This allows other concurrent local operations to make progress. 

\tpcw and \rubis show different results due to different read-only operation ratios. 
In \tpcw many of the local operations are write operations that, in \mysqlcluster, involve distributed transactions.
Therefore, \tpcw benefits tremendously from operation partitioning. 
The \rubis's workload contains more local operations, but a much larger fraction is read-only.
\rubis thus profits from the read-only transaction optimizations implemented by \mysqlcluster.
These results highlight that existing DBMSs already require minimal coordination for read-dominated workloads.
The more a workload is write-heavy, and thus hard to scale out, the more using \name pays off.

\subsection{RQ2: Scaling Out in WANs }
The previous experiments showed the scale-out capabilities of \name in a LAN setting.
We now evaluate \name in a WAN (i.e., geographically distributed) setting, where coordination is even more expensive and scalability is more challenging. 
We use two baselines:
\begin{inparaenum}[(1)]
\item a standard \mysql (without \name) a single server \textbf{(centralized)}, and 
\item an implementation where read-only operations are executed by one server without coordination, like local operations. This is a common optimization offered by many systems \textbf{(read-only)}.
\end{inparaenum}
All these variants guarantee serializability, so the applications have the impression of interacting with a single server and don't need to be modified to account for inconsistencies.

\begin{table}
\centering
\footnotesize
\begin{tabular}{|c||c|c|}
\hline
 Configuration &\tpcw\ & \rubis\ \\
 \hline
 \hline
 Centralized & 1390ms & 416ms \\
 \name -- 2 &  671ms (2.1x) & 122ms (3.4x)  \\
 \name -- 3 &  436ms (3.2x) & 155ms (2.7x)  \\
 \name -- 5 &  29ms (47.9x)& 35ms (11.9x)  \\
 Read-Only -- 2 &  902ms (1.5x) & 145ms (2.9x)  \\
 Read-Only -- 3 &  521ms (2.7x) & 131ms (3.2x)  \\
 Read-Only -- 5 &  129ms (10.8x) & 96ms (4.3x)  \\
 \hline
\end{tabular}
\caption{Request latency with light load in a WAN setting. The reported improvements in brackets are relative to the centralized case.}
\label{tab:latency}
\end{table}

\begin{figure}
	\centering
	\includegraphics[width=0.5\textwidth]{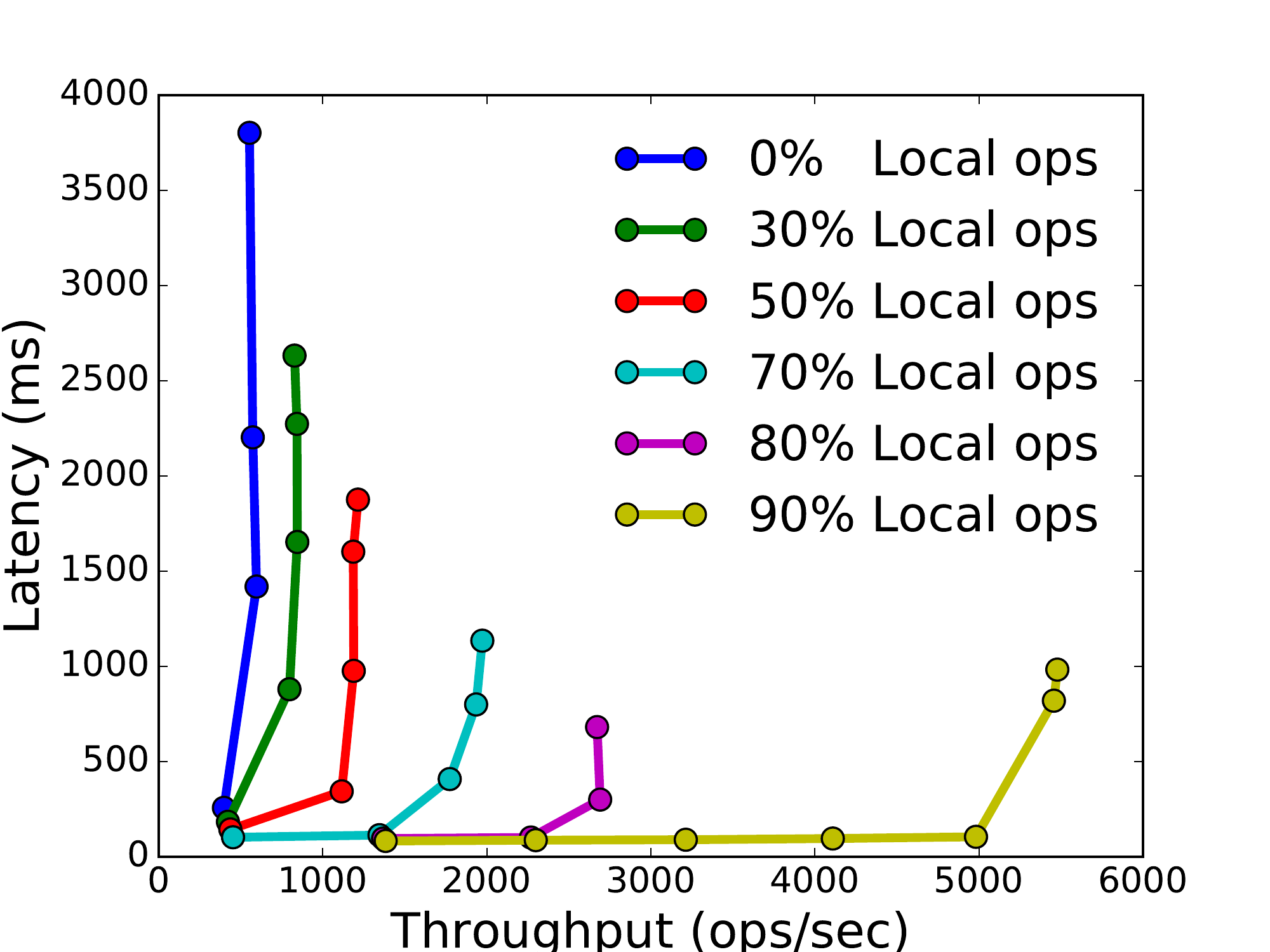}
	\caption{\name with different local operation ratios.}
	\label{fig:micro}
\end{figure}

First, we compare the latency of \NAME in different configurations when the system is not overloaded.
In Table~\ref{tab:latency}, we report the latency improvement over the centralized setting of each configuration, from two to five with \tpcw and \rubis. 

\name\ achieves significant latency reduction, of more than one order of magnitude, because it reduces the need for coordination.
For instance, \rubis with 3 servers the latency is 3.2x less that of a centralized server and 2.7 for the read-only baseline. The performance is best when a server datacenter is available in every geographical location of the clients. In fact, for the 5 server configuration the latency is 47.9x less for \tpcw and 11.9x for \rubis. In contrast, the latency when using the read-only optimizations is only 10.8x less for \tpcw and 4.3x for \rubis.
This is because the majority of operations can be served by the local server where clients are located.
This is especially the case for the five servers case where the latency is almost completely dominated by the time it takes to handle the queries locally. 
For instance, in Table \ref{tab:latency} we report the average latencies of \name in the five servers configurations which are in the order of 29ms for \rubis and 35ms for \tpcw. 
The intra-site latency (Table \ref{tab:ping}) is 20ms and constitutes about 69\% and 57\% of the latencies in \tpcw and \rubis, respectively.

Next, we shift our attention to both throughput and latency with more intense workloads (Figure~\ref{fig:WAN-experiment}).
We stress the system by increasing the number of clients until the latency reaches 5 seconds.
The single server in the centralized case start to saturate quickly, at few tens of operations per second.
Read-only optimization significantly reduces latency and increases throughput for both workloads and especially for \rubis which is more read-dominated.
\name, however, has a much larger impact as it allows the local execution of many more operations, both read-only and not. 
The effect in terms of throughput over both the centralized and read-only baselines is substantial. 
\name enables multiple sites to execute operations in parallel and results in much higher maximum throughput. 

Overall, \name improves the maximum throughput compared to the read-only setting. For instance, in the five servers configuration there is an increase of the maximal throughput by 291\% for \tpcw\ and 181\% for \rubis.
In terms of scalability, Figures~\ref{fig:tpcw-wan} and~\ref{fig:rubis-wan} show that \name scales very well until at least five geo-locations, which is a fairly high number in many practical settings. 
By contrast, the read-only baseline maxes out already with three servers, especially with \tpcw where the gain from using additional servers in terms of throughput is marginal.
Like in the LAN case, if we keep adding locations, we expect scalability to asymptotically stop because of the increasing coordination cost induced by executing global operations. 


\begin{figure}
	\centering
	\begin{subfigure}[b]{0.47\columnwidth}
		\includegraphics[width=1.1\textwidth]{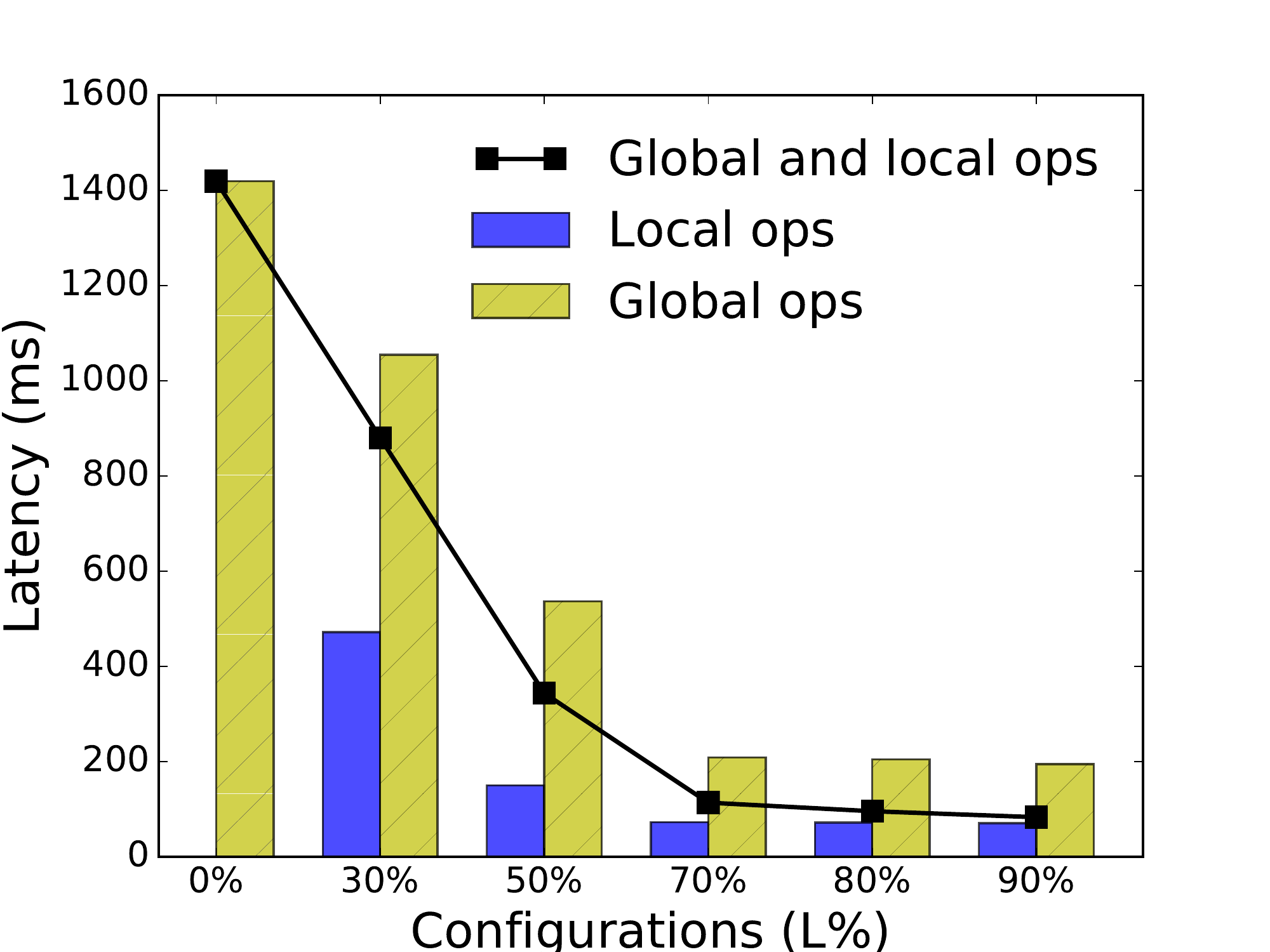}
		\caption{Light load.}
		\label{fig:micro-latencies}
	\end{subfigure}
	~
	\begin{subfigure}[b]{0.47\columnwidth}
		\includegraphics[width=1.1\textwidth]{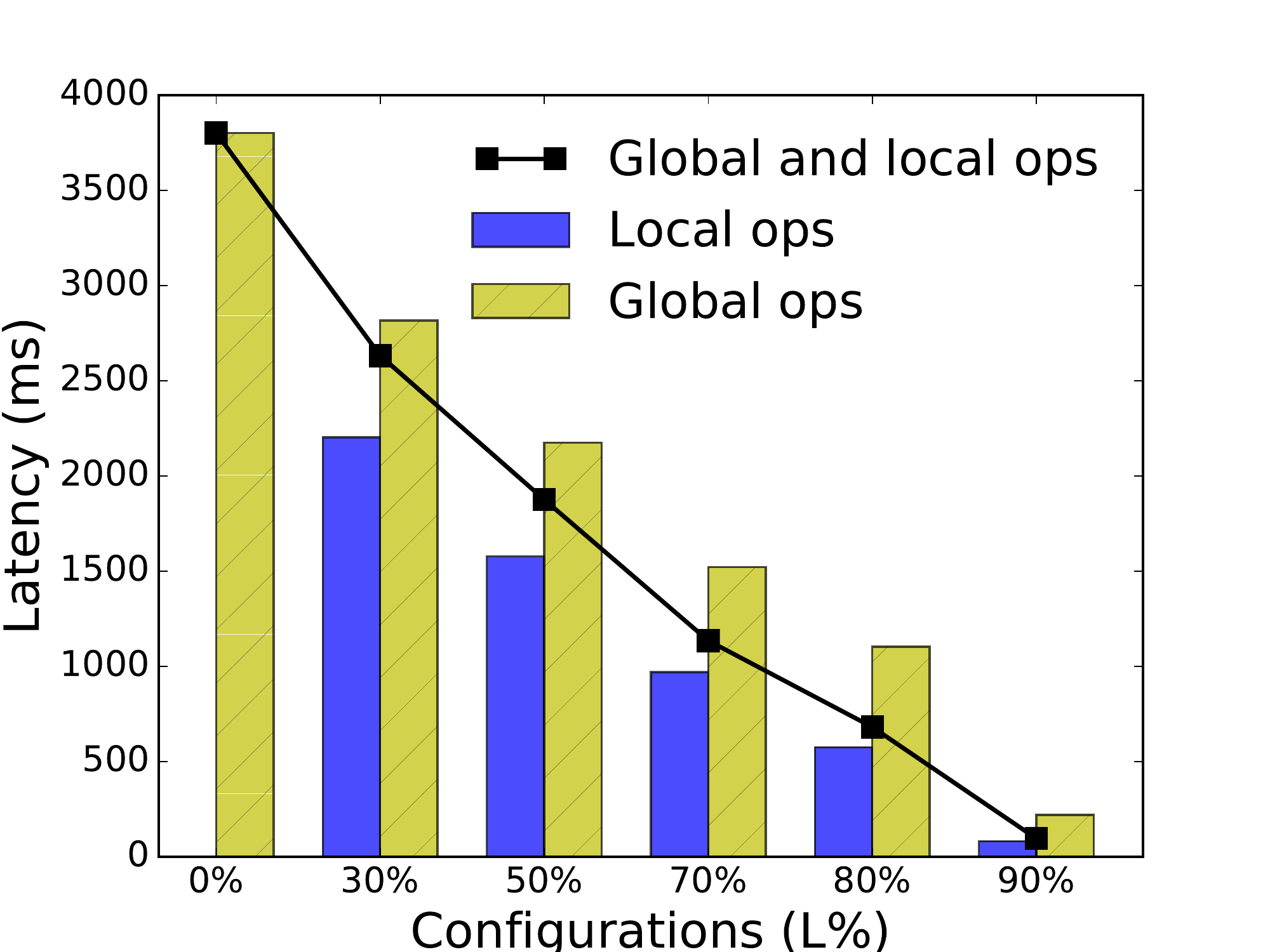}
		\caption{Heavy load.}
		\label{fig:micro-latencies-loaded}
	\end{subfigure}
	\caption{Latency comparison of \name with different local operation ratios using micro-benchmarks.}
	\label{fig:micro-bars}
\end{figure}

\subsection{RQ3: Micro-Benchmarks}

We now examine the performance of \name more in detail.
We analyze the effect of different local operations ratios on \name's performance using a synthetic workload where we can precisely specify these ratios. 
The execution time of operations (global or local) is fixed to 5ms. 
We use a WAN setup with three servers and vary the percentage of local operations in the workload from 0\% to 90\%.

Figure \ref{fig:micro} confirms that the performance of \name is highly sensitive to the fraction of local operations in the workload. 
For instance, with a workload of 30\% of local operations the system starts to saturate already around 600 ops/sec while in a workload of 90\% local operations the saturation starts only around the 5477 ops/sec. 
This can be explained by the additional coordination overhead of global operations unlike local operations which can be served by the nearest server.

Figure \ref{fig:micro-latencies} shows mean latencies for local and global operations with a light load (far from saturation). 
The average latency of all operations decreases as we add more local operations to the mix and he have less global operations queuing up.
As expected, in all configurations the latency of local operations is much lower and is between 2.23x and 3.75x less compared to that of global operations.
For instance, in a configuration with 70\% of local operations, the mean latency is 195ms for local operations and 70ms for global operations (2.78x less).
The overall latency stabilizes with 70\% local operations or more. 
By contrast, in a configuration with a higher load (Figure \ref{fig:micro-latencies-loaded}), the overall latency continues to fall even after the 70\% threshold observed in Figure \ref{fig:micro-latencies}. 
The reason is that the saturation of the system does not only occur because of the large fraction of global operations queuing up but also because of the overall volume of requests.


\section{Related Work}
\label{sec:background}

Scaling out client-server applications is an important topic and it has been the subject of a large volume of work.
We now review it and position the \NameAlgo approach in this landscape.
For space reasons, we do not review the literature on fault-tolerant replication algorithms since fault tolerance can be treated as an orthogonal issue to distributed transactions.
We leave combining the two problems, along the lines of work like~\cite{tapir}, as future work.

\spara{Data Partitioning}
The problem of finding an optimal database design is NP-hard~\cite{partition-np-hard-1}.
Nonetheless, a large number of heuristics for data partitioning have been proposed, such as~\cite{schism,horticulture,sword,jecb}. 
These techniques require substantial offline effort, including running a representative workload, collecting samples, defining an accurate cost model of the system performance, and sometimes user guidance in identifying the best solution.
\NameAlgo indirectly obtains a partial data partitioning scheme, much like existing work, but it is entirely automated and based on static analysis.

\spara{Distributed Transactions}
The typical approach to implement distributed transactions, which is used in many practical database management systems, is to lock the rows accessed by the transaction and to use two phase commit to conclude the transaction.
Since this approach is expensive, there has been much work on speeding up distributed transactions.
Spanner speeds up read-only transactions through the use of synchronized clocks~\cite{spanner}.
H-Store speeds up ACID transactions that access only a single partition.
It supports multi-partition transactions using standard locking and two-phase commit protocols.
Our evaluation shows that the \NameProt is superior to a standard two phase commit transnational system with locks.
ElasTraS~\cite{elastras}, G-Store~\cite{gstore}, and MegaStore~\cite{megastore} only support ACID transactions within the boundary of a single partition or key group, and do not offer full transactional support like \name.

Several approaches like Calvin~\cite{calvin}, Lynx~\cite{lynx}, Rococo~\cite{rococo}, Callas~\cite{callas} and others~\cite{faleiro2015rethinking,transaction-chopping,callas2}, have been proposed to improve the performance of distributed transactions, but they typically require implementing a novel database management or data store system, unlike the \NameProt protocol which is a middleware running on top of an unmodified black-box, single-server DBMS offering a JDBC interface.
In addition, they require additional knowledge about the semantic of the application that must be provided by the user, sometimes by restructuring the code.
Modifying and extending the application code in this sense can be complex and cumbersome, and sometimes unfeasible.
The \NameProt protocol does not have this requirement, since \NameAlgo applies to unmodified application code.
Yet, \NameProt provides competitive performance speedups.
While the Callas algorithm supports serializability, the actual Callas prototype system only provides the {\em read committed} isolation level, just like the \mysqlcluster system it is based upon.
\name provides serializability instead, which is significantly more expensive than read committed isolation.
Nonetheless, \name achieves similar speedups over \mysqlcluster as the Callas results reported in~\cite{callas}.


SDD-1~\cite{sdd1} is related to our approach in that it uses {\em transaction classes}, but still differs in several aspects.
First, a key pre-step to achieve good performance in SDD-1 is that the user provides a good grouping of transaction into classes, but SDD-1 offers no support for it.
In our approach, we automatically generate operation partitions that can be leveraged by our protocol based on static analysis. 
Second, SDD-1 replicas executing global operations need blocking coordination based on timestamps.
This algorithm was pioneering work on distributed transactions, but is less efficient than algorithms based on distributed locks~\cite{phil-fads}, which we compare against.

\spara{Weakly consistent scale-out approaches}
Most algorithms using replication to scale out offer only weak consistency guarantees: eventual consistency~\cite{bayou,dynamo} session consistency~\cite{session}, causal consistency~\cite{cops}, timeline consistency~\cite{pnuts}, and Parallel Snapshot Isolation~\cite{psi}.

Recent work proposes strengthening weak consistency with {\em invariants}, like in the Red/Blue model~\cite{redblue}, the Explicit Consistency model~\cite{putting}, and Invariance Confluence~\cite{coordination-avoidance}.
Requiring developers to define good invariants is challenging.
Also, even with invariants, the system will still show a weakly-consistent behavior that would not occur in a sequential execution.
Unlike these approaches, \NameAlgo support serializability~\cite{serializability}, as required by ACID applications.

\spara{Treaties}
Prior work on treaties combines scale-out replication and strong consistency for subset of operations.
Informally, treaties allow replicas to agree to split the value of a certain field and to share the splits.
For example, in a ticket sale application, replicas can agree on a treaty where each take a share of the available tickets, so that they do not need to coordinate every time they sell a ticket unless they sell out their share.
Treaties make specific assumptions on the applications they target: concurrent transactions must make small commutative modifications to a shared global quantity at different replicas, and their outcome must not be sensitive to such small modifications.
Examples of treaties are the escrow protocol~\cite{escrow}, the demarcation protocol~\cite{demarcation}, Homeostasis~\cite{homeostasis}, and time-limited warranties~\cite{warranties}.
Work related to the idea of treaties has also investigated relaxed notions of consistency such as bounded inconsistency~\cite{vahdat} or consistency rationing~\cite{rationing}.
\NameAlgo is more generic as it does not make assumptions on the application, as treaties do.
\NameAlgo can be applied to any application, whereas treaties require either user knowledge about the application semantic or the use of special languages, like in Homeostasis.




\section{Conclusion}
We introduced \NameAlgo, a technique that allows scaling out applications while preserving serializability.
We implement our technique in a middleware, called \NAME that can be used with an unmodified DBMS.
Our experiments with two user application \tpcw and \rubis show that \NAME is very effective in both LAN and WAN settings.
\balance


{
\bibliography{references}}

\begin{thebibliography}{10}

\bibitem{jparser}
Java parser.
\newblock https://github.com/javaparser/javaparser, 2015.

\bibitem{ebay}
Ebay website.
\newblock https://www.ebay.com, 2017.

\bibitem{coordination-avoidance}
P.~Bailis, A.~Fekete, M.~J. Franklin, A.~Ghodsi, J.~M. Hellerstein, and
  I.~Stoica.
\newblock Coordination avoidance in database systems.
\newblock {\em Proceedings of the VLDB Endowment}, 8(3):185--196, 2014.

\bibitem{megastore}
J.~Baker, C.~Bond, J.~C. Corbett, J.~Furman, A.~Khorlin, J.~Larson, J.-M. Leon,
  Y.~Li, A.~Lloyd, and V.~Yushprakh.
\newblock Megastore: Providing scalable, highly available storage for
  interactive services.
\newblock In {\em CIDR}, volume~11, pages 223--234, 2011.

\bibitem{putting}
V.~Balegas, S.~Duarte, C.~Ferreira, R.~Rodrigues, N.~Pregui{\c{c}}a,
  M.~Najafzadeh, and M.~Shapiro.
\newblock Putting consistency back into eventual consistency.
\newblock In {\em Proceedings of the Tenth European Conference on Computer
  Systems}, page~6. ACM, 2015.

\bibitem{demarcation}
D.~Barbard and H.~Garcia-Molina.
\newblock The demarcation protocol: A technique for maintaining linear
  arithmetic constraints in distributed database systems.
\newblock In {\em Advances in Database Technology - EDBT'92}, pages 373--388.
  Springer, 1992.

\bibitem{phil-fads}
P.~Bernstein.
\newblock Timestamp-based concurrency control in sdd-1.
\newblock In {\em Workshop in Failed Aspirations in Database Systems}, 2017.

\bibitem{sdd1}
P.~A. Bernstein, J.~Rothnie, N.~Goodman, and C.~A. Papadimitriou.
\newblock The concurrency control mechanism of sdd-1: A system for distributed
  databases (the fully redundant case).
\newblock {\em IEEE Transactions on Software Engineering}, 4(3):154, 1978.

\bibitem{tpcw}
T.~Consortium.
\newblock {TPC} benchmark-{W} specification v. 1.8.
\newblock http://www.tpc.org/tpcw/spec/tpcw\_v1.8.pdf, 2002.

\bibitem{pnuts}
B.~F. Cooper, R.~Ramakrishnan, U.~Srivastava, A.~Silberstein, P.~Bohannon,
  H.-A. Jacobsen, N.~Puz, D.~Weaver, and R.~Yerneni.
\newblock Pnuts: Yahoo!'s hosted data serving platform.
\newblock {\em Proceedings of the VLDB Endowment}, 1(2):1277--1288, 2008.

\bibitem{spanner}
J.~C. Corbett, J.~Dean, M.~Epstein, A.~Fikes, C.~Frost, J.~J. Furman,
  S.~Ghemawat, A.~Gubarev, C.~Heiser, P.~Hochschild, et~al.
\newblock Spanner: Google's globally distributed database.
\newblock {\em ACM Transactions on Computer Systems (TOCS)}, 31(3):8, 2013.

\bibitem{schism}
C.~Curino, E.~Jones, Y.~Zhang, and S.~Madden.
\newblock {Schism: A Workload-driven Approach to Database Replication and
  Partitioning}.
\newblock {\em PVLDB}, 3(1-2), 2010.

\bibitem{gstore}
S.~Das, D.~Agrawal, and A.~El~Abbadi.
\newblock G-store: a scalable data store for transactional multi key access in
  the cloud.
\newblock In {\em Proceedings of the 1st ACM symposium on Cloud computing},
  pages 163--174. ACM, 2010.

\bibitem{elastras}
S.~Das, A.~El~Abbadi, and D.~Agrawal.
\newblock Elastras: An elastic transactional data store in the cloud.
\newblock {\em HotCloud}, 9:131--142, 2009.

\bibitem{dynamo}
G.~DeCandia, D.~Hastorun, M.~Jampani, G.~Kakulapati, A.~Lakshman, A.~Pilchin,
  S.~Sivasubramanian, P.~Vosshall, and W.~Vogels.
\newblock Dynamo: amazon's highly available key-value store.
\newblock In {\em ACM SIGOPS Operating Systems Review}, volume~41, pages
  205--220. ACM, 2007.

\bibitem{rubis}
C.~Emmanuel and M.~Julie.
\newblock Rubis: {R}ice university bidding system.
\newblock http://rubis.ow2.org/.

\bibitem{faleiro2015rethinking}
J.~M. Faleiro and D.~J. Abadi.
\newblock Rethinking serializable multiversion concurrency control.
\newblock {\em Proceedings of the VLDB Endowment}, 8(11):1190--1201, 2015.

\bibitem{zab}
F.~P. Junqueira, B.~C. Reed, and M.~Serafini.
\newblock Zab: High-performance broadcast for primary-backup systems.
\newblock In {\em Dependable Systems \& Networks (DSN), 2011 IEEE/IFIP 41st
  International Conference on}, pages 245--256. IEEE, 2011.

\bibitem{difference}
F.~P. Junqueira and M.~Serafini.
\newblock On barriers and the gap between active and passive replication.
\newblock In {\em Distributed Computing}, pages 299--313. Springer, 2013.

\bibitem{rationing}
T.~Kraska, M.~Hentschel, G.~Alonso, and D.~Kossmann.
\newblock Consistency rationing in the cloud: pay only when it matters.
\newblock {\em Proceedings of the VLDB Endowment}, 2(1):253--264, 2009.

\bibitem{redblue}
C.~Li, D.~Porto, A.~Clement, J.~Gehrke, N.~Pregui{\c{c}}a, and R.~Rodrigues.
\newblock Making geo-replicated systems fast as possible, consistent when
  necessary.
\newblock In {\em Presented as part of the 10th USENIX Symposium on Operating
  Systems Design and Implementation (OSDI 12)}, pages 265--278, 2012.

\bibitem{warranties}
J.~Liu, T.~Magrino, O.~Arden, M.~D. George, and A.~C. Myers.
\newblock Warranties for faster strong consistency.
\newblock In {\em 11th USENIX Symposium on Networked Systems Design and
  Implementation (NSDI 14)}, pages 503--517, 2014.

\bibitem{cops}
W.~Lloyd, M.~J. Freedman, M.~Kaminsky, and D.~G. Andersen.
\newblock Don't settle for eventual: scalable causal consistency for wide-area
  storage with cops.
\newblock In {\em Proceedings of the Twenty-Third ACM Symposium on Operating
  Systems Principles}, pages 401--416. ACM, 2011.

\bibitem{rococo}
S.~Mu, Y.~Cui, Y.~Zhang, W.~Lloyd, and J.~Li.
\newblock Extracting more concurrency from distributed transactions.
\newblock In {\em 11th USENIX Symposium on Operating Systems Design and
  Implementation (OSDI 14)}, pages 479--494, 2014.

\bibitem{partition-np-hard-1}
R.~Mukkamala, S.~C. Bruell, and R.~K. Shultz.
\newblock {\em Design of partially replicated distributed database systems: an
  integrated methodology}, volume~16.
\newblock ACM, 1988.

\bibitem{escrow}
P.~E. O'Neil.
\newblock The escrow transactional method.
\newblock {\em ACM Transactions on Database Systems (TODS)}, 11(4):405--430,
  1986.

\bibitem{serializability}
C.~H. Papadimitriou.
\newblock The serializability of concurrent database updates.
\newblock {\em Journal of the ACM (JACM)}, 26(4):631--653, 1979.

\bibitem{horticulture}
A.~Pavlo, C.~Curino, and S.~Zdonik.
\newblock Skew-aware automatic database partitioning in shared-nothing,
  parallel oltp systems.
\newblock In {\em Proceedings of the 2012 ACM SIGMOD International Conference
  on Management of Data}, pages 61--72. ACM, 2012.

\bibitem{bayou}
K.~Petersen, M.~Spreitzer, D.~Terry, and M.~Theimer.
\newblock Bayou: replicated database services for world-wide applications.
\newblock In {\em Proceedings of the 7th workshop on ACM SIGOPS European
  workshop: Systems support for worldwide applications}, pages 275--280. ACM,
  1996.

\bibitem{sword}
A.~Quamar, K.~A. Kumar, and A.~Deshpande.
\newblock Sword: scalable workload-aware data placement for transactional
  workloads.
\newblock In {\em International Conference on Extending Database Technology},
  pages 430--441, 2013.

\bibitem{homeostasis}
S.~Roy, L.~Kot, G.~Bender, B.~Ding, H.~Hojjat, C.~Koch, N.~Foster, and
  J.~Gehrke.
\newblock The homeostasis protocol: Avoiding transaction coordination through
  program analysis.
\newblock In {\em Proceedings of the 2015 ACM SIGMOD International Conference
  on Management of Data}, pages 1311--1326. ACM, 2015.

\bibitem{transaction-chopping}
D.~Shasha, F.~Llirbat, E.~Simon, and P.~Valduriez.
\newblock Transaction chopping: Algorithms and performance studies.
\newblock {\em ACM Transactions on Database Systems (TODS)}, 20(3):325--363,
  1995.

\bibitem{psi}
Y.~Sovran, R.~Power, M.~K. Aguilera, and J.~Li.
\newblock Transactional storage for geo-replicated systems.
\newblock In {\em Proceedings of the Twenty-Third ACM Symposium on Operating
  Systems Principles}, pages 385--400. ACM, 2011.

\bibitem{callas2}
C.~Su, N.~Crooks, C.~Ding, L.~Alvisi, and C.~Xie.
\newblock Bringing modular concurrency control to the next level.
\newblock In {\em Proceedings of the 2017 ACM International Conference on
  Management of Data}, SIGMOD '17, pages 283--297, New York, NY, USA, 2017.
  ACM.

\bibitem{session}
D.~B. Terry, A.~J. Demers, K.~Petersen, M.~J. Spreitzer, M.~M. Theimer, and
  B.~B. Welch.
\newblock Session guarantees for weakly consistent replicated data.
\newblock In {\em Parallel and Distributed Information Systems, 1994.,
  Proceedings of the Third International Conference on}, pages 140--149. IEEE,
  1994.

\bibitem{calvin}
A.~Thomson, T.~Diamond, S.-C. Weng, K.~Ren, P.~Shao, and D.~J. Abadi.
\newblock Calvin: fast distributed transactions for partitioned database
  systems.
\newblock In {\em Proceedings of the 2012 ACM SIGMOD International Conference
  on Management of Data}, pages 1--12. ACM, 2012.

\bibitem{jecb}
K.~Q. Tran, J.~F. Naughton, B.~Sundarmurthy, and D.~Tsirogiannis.
\newblock {JECB}: A join-extension, code-based approach to {OLTP} data
  partitioning.
\newblock In {\em ACM SIGMOD International Conference on Management of Data},
  pages 39--50, 2014.

\bibitem{callas}
C.~Xie, C.~Su, C.~Littley, L.~Alvisi, M.~Kapritsos, and Y.~Wang.
\newblock High-performance acid via modular concurrency control.
\newblock In {\em Proceedings of the 25th Symposium on Operating Systems
  Principles}, pages 279--294. ACM, 2015.

\bibitem{vahdat}
H.~Yu and A.~Vahdat.
\newblock Design and evaluation of a continuous consistency model for
  replicated services.
\newblock In {\em Proceedings of the 4th conference on Symposium on Operating
  System Design \& Implementation-Volume 4}, pages 21--21. USENIX Association,
  2000.

\bibitem{tapir}
I.~Zhang, N.~K. Sharma, A.~Szekeres, A.~Krishnamurthy, and D.~R. Ports.
\newblock Building consistent transactions with inconsistent replication.
\newblock In {\em Proceedings of the 25th Symposium on Operating Systems
  Principles}, pages 263--278. ACM, 2015.

\bibitem{lynx}
Y.~Zhang, R.~Power, S.~Zhou, Y.~Sovran, M.~K. Aguilera, and J.~Li.
\newblock Transaction chains: achieving serializability with low latency in
  geo-distributed storage systems.
\newblock In {\em Proceedings of the Twenty-Fourth ACM Symposium on Operating
  Systems Principles}, pages 276--291. ACM, 2013.

\end{thebibliography}
\newpage

\section*{Appendix: Correctness Proof}

In this section, we prove that the \NameProt protocol (Algorithm~\ref{alg:coord-passive}) correctly implements serializability.

\subsection*{Token-Passing Scheme}

The protocol uses a primary-backup scheme to execute global operations, based on a token passing scheme that acts as a broadcast algorithm.
We now show the properties of the token passing algorithm.

\begin{lemma}
\label{theo:poabcast}
The token passing scheme used by the \NameProt protocol to broadcast state updates of global operations satisfies Primary Order atomic broadcast~\cite{zab,difference}.
\end{lemma}

\textbf{Proof.} A Primary Order atomic broadcast protocol satisfies the standard properties of atomic broadcast, namely:
\begin{compactitem}
\item {\em integrity:} if some server delivers an update $u$ then some server has appended $u$ to the token; 
\item {\em total order:} if some server delivers update $u$ before $u'$ then any server that delivers $u'$ must deliver $u$ before $u'$; 
\item {\em agreement:} if some server $p$ deliver $u$ and some other server $q$ delivers $u'$, then either $p$ delivers $u'$ or $q$ delivers $u$.
\end{compactitem} 

These properties holds for the token scheme since the order of the updates appended to the token is never altered and updates are only removed from the token once all servers has received them (Line \ref{ln:token-remove}).

There are two additional properties that Primary Order atomic broadcast satisfies. 
The first additional property is {\em primary order:} servers must apply updates in the order in which they were broadcast by the primaries that produced them.
This is necessary because otherwise older state updates might overwrite values written by newer state updates. 
In particular, {\em local} primary order requires that the delivery order is consistent with the local broadcast order of a primary during each primary epoch, while {\em global} primary order requires that the delivery order is consistent with the total order of the primary epochs in which the message was broadcasted.

The second property, called {\em primary integrity}, guarantees that the primary role can transition safely from one server $p$ to another server $q$.
Primary integrity requires the following: if a new primary epoch $e$ starts at server $p$, a state update $u$ is broadcasted during a prior primary epoch, and $u$ is eventually delivered by some server, then $p$ must deliver $u$ before it starts $e$.
This property guarantees that the new primary $p$ obtains the full final state resulting from previous epoch {\em before} it starts producing new state updates.
State updates are incremental and they should only be applied on the state from which they were produced.
Interleaving state updates from different epochs can result in incorrect executions.

It is easy to see that the token passing scheme satisfies these properties.
Primary order is guaranteed because the updates are appended to the token in the same order of their execution by the primary (Lines \ref{ln:q-execute}-\ref{ln:q-append}) and because of the total order preservation of the token.
This is true because the used queue is atomic 
Updates appended to the token are applied by all the other servers in that same order(Lines \ref{ln:token-start}-\ref{ln:token-end}).
The token scheme satisfies primary integrity by ensuring that every primary applies all the updates in the token from previous epochs before executing pending global operations (Lines \ref{ln:token-start}-\ref{ln:token-end}).$\hfill \Box$

\BlankLine

\subsection*{Serializability Proof}

We now show that \NameProt guarantees serializability, that is, the relative order of global operation is consistent across all servers.
Before the proof, we need to introduce some notation and definitions.

\spara{Definitions}
In the \NameProt protocol, operations are executed concurrent by multiple threads.
However, they are executed by an underlying DBMS which, by assumption, serializes their execution.
Therefore, we consider in the proof that each server executes operations in a sequential total order.
A DBMS running at a server only executes state updates of global operation of other servers.
In the proof, we will not distinguish between the two cases of executing a global operation or its state update, that is, we say that a server executes a global operation  $g$ also when it executes the corresponding state update.

We call an {\em execution} of the system the sequence of operations invoked on the distributed system up to a given time, and pair each operation with the reply that the service produced for it.
In order to show that the \NameProt protocol satisfies {\em serializability}~\cite{serializability}, we need to show that, for each possible execution $e$, there exists a total order $T$ of the operation-reply pairs of $e$ such that executing of the operations in the specified order on a single instance of the service will produce the same replies as in $e$.

It is important to stress that this total execution order $T$ is a {\em logical} order.
It describes how clients observe the behavior of an application scaled out by a coordination protocol like the \NameProt protocol. 
The actual implementation of the coordination protocol simply has to exhibit a behavior that is {\em equivalent} to this total order.
The implementation of this property is protocol-dependent.
In the \NameProt protocol, for example, local operations are not executed by all servers.
Nonetheless, local operations are still totally ordered in $T$, and the system behaves {\em as if} these operations were executed by all servers.

We now introduce additional notation.
Let $T'_p$ be the execution order of all global operations executed in $e$ relative to server $p$.
$T'_p$ is defined as follows. 
Let $g$ be a global operation that appears in $e$.
If $g \in G_p$, then $T'_p$ orders $g$ according to the order in which $g$ is executed at $p$.
Else, $T'_p$ orders $g$ according to the order in which the state update generated from $g$ is applied at $p$.
Since in each server the token thread executes and applies global operations sequentially, $T'_p$ is a total order, and it reflects the order in which global operations modify the state of $p$.
Note that the total order for two servers might contain a different set of operations.
For simplicity, we treat sometimes $T'_p$ as a set and use the notation $g \in T'_p$ to say that operation $g$ appears in the total order $T'_p$.

\spara{Serializability proof}
The proof is in three steps.
First, we show that each server orders pairs of global operations consistently.
Next, we show that the relative execution order of local and global operations is consistent across all servers.
Finally, we show that pairs of local operations are executed consistently.

\begin{algorithm}[t]
\caption{The \NameProt algorithm for server $p$  (copied from Algorithm~\ref{alg:coord-passive})}
\label{alg:coord-passive-appendix}
\begin{footnotesize}

\Event receive $\langle \textrm{REQ}, o, c \rangle$ msg from client $c$ where \nllabel{ln:c-begin}\Do{
	\If{$o \in C \cup L_p$\nllabel{ln:cl-begin}}{
		$r, *\leftarrow$ execute($o$)\;
		send $\langle \textrm{REPLY}, r \rangle$ msg to $c$\nllabel{ln:cl-end}\;	
	} \ElseIf{$o \in G_p$\nllabel{ln:g-begin}}{
		append $\langle o,c\rangle$ to $Q$\nllabel{ln:g-end}\;
	} \Else {
		$q \leftarrow$ replica such that $o \in L_q \cup G_q$\nllabel{ln:map-begin}\; 
		send $\langle \textrm{MAP}, q \rangle$ msg to $c$\nllabel{ln:map-end}\;
	}
}

\BlankLine

\Event event \textsc{receiveToken}$(T)$ \nllabel{ln:primary-begin}\Do{
	\ForEach{$\langle u,q \rangle \in T$\nllabel{ln:token-start}}{
		\If{$p = q$}{
      remove $\langle u,q \rangle$ from $T$\nllabel{ln:token-remove}\;
		}
		\Else{
			apply($u$)\nllabel{ln:token-end}\;
		}
	}
	$Q' \leftarrow $ atomic-snapshot($Q$)\nllabel{ln:copy}\;
	\ForEach{$\langle o,c \rangle \in Q'$\nllabel{ln:q-start}}{
		$r, u \leftarrow$ execute($o$)\nllabel{ln:q-execute}\;
    append $\langle u, p \rangle$ to $T$\nllabel{ln:q-append}\;
		send $\langle \textrm{REPLY}, r \rangle$ msg to $c$\;
		remove $\langle o,c \rangle$ from $Q$\nllabel{ln:q-end}\;
	}
	\textsc{passToken}$(T)$\nllabel{ln:token-release}\;
} 
\end{footnotesize}
\end{algorithm}

\begin{lemma}
\label{theo:global}
Given two servers $p$ and $q$, their total orders $T'_p$ and $T'_q$ have a common prefix which includes every operation in $T'_p \cap T'_q$.
\end{lemma}

\textbf{Proof.} 
This lemma directly follows from the fact that the state updates of global operations are broadcasted and delivered using the token scheme.
The token scheme guarantees that the delivery order of state updates is consistent across all servers.
, so the pairwise order of global operations in $T'_p \cap T'_q$ that are neither in $G_p$ nor in $G_q$ is consistent in $T'_p$ and $T'_q$, and we are done.

Consider now the ordering of two global operations $g$ and $g'$ such that $g \in G_p$ or $g \in G_q$ (remember that global operations are partitioned).
We consider only the case $g \in G_p$ without loss of generality, and we have two sub-cases:

{\em Case I:} If $g' \in G_p$, then the primary order property of the token scheme guarantees that the delivery order of the state updates of $g$ and $g'$ at $q$ is consistent with the order in which the operations were executed by $p$, and we are done.

{\em Case II:} If $g' \not\in G_p$, we consider two sub-cases:

{\em Case II.a}: $g'$ precedes $g$ in $T'_p$. 
The primary integrity property guarantees that, before $p$ becomes a primary and starts executing new operations, $p$ also delivers all operations sent by previous primaries that are ever delivered by $q$.
Therefore, $g'$ is delivered at $p$ before $g$ is executed, so the same order will appear in $T'_q$.

{\em Case II.b}: $g'$ follows $g$ in $T'_p$. 
There exists some server $r$ such that $g' \in G_r$. 
If $r$ sends $g'$ in a primary epoch after $p$ sends $g$, then $r$ delivers $g$ before it executes and sends $g'$ by primary integrity.
Therefore, server $r$ delivers $g$ before $g'$.
Because of the total order guaranteed by the token scheme, every server must deliver $g$ before $g'$, including $p$.
This implies that $g$ must precede $g'$ in $T'_p$, which is a contradiction.

If $r$ sends $g'$ in a primary epoch before $p$ sends $g$, then $g'$ precedes $g$ in $T'_p$ by primary integrity.
It follows that $g'$ precedes $g$ in $T'_q$ too.
Assume by contradiction that this does not hold and $g$ precedes $g'$ in $T'_q$.
This would imply that $q$ delivers the state update for $g$ before the one for $g'$.
Because of the total order property of the token scheme, also $p$ should deliver the state updates in the same order, so $p$ should deliver the state update for $g$ before producing it, a contradiction.$\hfill \Box$

\BlankLine

Since all servers order global operations consistently, we define the order of global operations in the total order $T$ according to the order $T'_p$ of any server $p$.

Next, we can show how pairs of operations, one local and one global, are ordered among each other.
Let $l \in L_p$ be a local operation executed by server $p$.
Let $B_p^l$ (resp. $A_p^l$) be the set of global operations whose state updates have been delivered at $p$ before (resp. after) $p$ executes $l$.
The total order $T$ orders $l$ after all global operations in $B_p^l$ and before the global operations in $A_p^l$. 

For this order to be sound, we need to show that $l$ the state updates of all global operations in $B_p^l$ are reflected in the state upon which $l$ is executed, and that the state update of $l$ is reflected in the state upon which all global operations in $A_p^l$ are executed.
The first claim directly follows from the fact that $l$ is executed by $p$ after the operations in $B_p^l$.
We now show the second claim.

\begin{lemma}
\label{theo:local}
The state update generated by executing $l$ at server $p$ is applied to the state upon which each global operation $g \in A_p^l$ is executed.
\end{lemma}

\textbf{Proof.} 
We consider two cases.
{\em Case I:} If $g \in G_p$, then $g$ is executed at $p$ and after $l$, by definition, so the state updated of $l$ is reflected in the local state of $p$ upon which $g$ is executed, and we are done.
{\em Case II:} If $g \in G_q$ with $q \neq p$, then $g$ does not directly read any variable from $l$ by definition (see Section~\ref{sec:classes}).
However, assume that, if operations were executed in the total order $T$, the state update of $l$ would determine the state upon which $g$ is executed through one (or more) operation $o$ that reads a value $v$ written by $l$ and, because of reading $v$, writes some value read by $g$. 
We need to show that these operations are actually executed before $g$.
We consider the case where there is only one operation $o$ propagating the changes of $l$ to $g$.
If $o$ reads from $l$ then $o$ is an operation assigned to server $p$.
If $g$, which is assigned to server $q$, reads from an operation $o$ at another server $p$, then $o$ is a global operation.
As shown in Lemma~\ref{theo:global} that global operations are consistently ordered by all servers.
We have also shown that this order reflects the execution order of global operations, so if $o$ precedes $g$ in the total order $T$, then the state update of $o$ takes effect before $g$ is executed.
The case where $l$ influences a global operation in $A_p^l$ through a chain of operations $o_1, \ldots, o_n$ follows by induction using a similar argument for each pair of subsequent operations in the chain.$\hfill \Box$

\BlankLine

The last case to consider is the ordering of pairs of local operations $l_1, l_2$.
If $l_1, l_2 \in L_p$ are executed by same server $p$, their correct order follows the local execution order at $p$.
The ordering between two local operations at different sites can be arbitrary, since neither observes the other by definition (see Section~\ref{sec:classification}).
Commutative operations can also be ordered arbitrarily with respect to any other operation in $T$.

After showing that the pairwise order of all operations form a consistent total order, and that this total order is consistent with the execution order of the operations, we can conclude that:

\begin{theorem}
\label{theo:final}
The \NameProt protocol (Algorithm~\ref{alg:coord-passive}) satisfies serializability.
\end{theorem}

\end{document}